\documentclass[a4paper,fleqn]{cas-dc}

\usepackage[numbers]{natbib} 
\usepackage{multirow}
\usepackage{listings}
\usepackage{subcaption}
\usepackage[T1]{fontenc}
\usepackage{color}
\usepackage{tikz}
\usepackage{xspace}
\usepackage{xcolor}
\usepackage{flushend}

\usetikzlibrary{positioning}

\def\BibTeX{{\rm B\kern-.05em{\sc i\kern-.025em b}\kern-.08emT\kern-.1667em\lower.7ex\hbox{E}\kern-.125emX}}

\newcommand{\bachelor}{\textsf{BSc}\xspace}
\newcommand{\master}{\textsf{MSc}\xspace}

\newcommand{\TUD}{\textbf{TUD}\xspace}
\newcommand{\TUe}{\textbf{TU/e}\xspace}
\newcommand{\OU}{\textbf{OU}\xspace}
\newcommand{\RU}{\textbf{RU}\xspace}
\newcommand{\RUG}{\textbf{RUG}\xspace}
\newcommand{\UvA}{\textbf{UvA}\xspace}
\newcommand{\UL}{\textbf{UL}\xspace}
\newcommand{\UT}{\textbf{UT}\xspace}
\newcommand{\UU}{\textbf{UU}\xspace}
\newcommand{\VU}{\textbf{VU}\xspace}
\newcommand{\SE}{SE\xspace}

\newcommand{\ka}[1]{\textsf{#1}\xspace}
\newcommand{\req}{\ka{Requirements}}
\newcommand{\arch}{\ka{Architecture}}
\newcommand{\des}{\ka{Design}}
\newcommand{\cnp}{\ka{Construction \& Programming}}
\newcommand{\test}{\ka{Testing}}
\newcommand{\ops}{\ka{SE Operations}}
\newcommand{\maint}{\ka{Maintenance}}
\newcommand{\conf}{\ka{Configuration Management}}
\newcommand{\semgmt}{\ka{SE management}}
\newcommand{\seproc}{\ka{SE process}}
\newcommand{\modeling}{\ka{SE models}}
\newcommand{\security}{\ka{Software Security}}
\newcommand{\economics}{\ka{SE economics}}
\newcommand{\verif}{\ka{Verification}}
\newcommand{\pldesign}{\ka{PL Design}}

\newcommand{\seminar}{\ka{Seminar}}
\newcommand{\prog}{\ka{Programming}}
\newcommand{\senum}{\ka{SE101}}
\newcommand{\proj}{\ka{Project}}
\newcommand{\ReqEng}{\emph{Requirements Engineering}}
\newcommand{\SSpe}{\emph{Software Specification}}
\newcommand{\BIS}{\emph{Business Information Systems}}
\newcommand{\FAS}{\emph{Fundamentals of Adaptive Software}}
\newcommand{\SBIS}{\emph{Seminar Business Information Systems}}
\newcommand{\ICTA}{\emph{ICT Architectures}}
\newcommand{\AIS}{\emph{Architectures of Information Systems}}
\newcommand{\DA}{\emph{Digital Architecture}}
\newcommand{\SD}{\emph{Software Design}}
\newcommand{\PA}{\emph{Problem Analysis}}
\newcommand{\DC}{\emph{Design for Change}}
\newcommand{\SDE}{\emph{System Design Engineering}}
\newcommand{\SOD}{\emph{Service Oriented Design}}
\newcommand{\SSSec}{\emph{System and Software Security}}
\newcommand{\CE}{\emph{Cryptographic Engineering}}
\newcommand{\PST}{\emph{Project System Testing}}
\newcommand{\SC}{\emph{Software Containerisation}}
\newcommand{\DevCS}{\emph{DevOps and Cloud-Based Systems}}
\newcommand{\SEM}{\emph{Software Engineering Methods}}
\newcommand{\SP}{\emph{Software Process}}
\newcommand{\SAM}{\emph{Software Asset Management}}
\newcommand{\SDM}{\emph{Systems Development Management}}
\newcommand{\SPM}{\emph{Software Project Management}}
\newcommand{\SM}{\emph{Software Management}}
\newcommand{\PM}{\emph{Project Management}}
\newcommand{\MDD}{\emph{Model-Driven Development}}
\newcommand{\MDE}{\emph{Model-Driven Engineering}}
\newcommand{\MDCPS}{\emph{Model-Based Design of Cyber-Physical Systems}}
\newcommand{\secure}{\emph{Security}}
\newcommand{\SV}{\emph{Security Verification}}

\usepackage{rotating}
\usepackage{adjustbox}

\definecolor{vstrong_pos}{HTML}{006837}
\definecolor{strong_pos}{HTML}{31a354}
\definecolor{moderate_pos}{HTML}{78C679}
\definecolor{weak_pos}{HTML}{C2E699}

\definecolor{vstrong_neg}{HTML}{a50f15}
\definecolor{strong_neg}{HTML}{de2d26}
\definecolor{moderate_neg}{HTML}{fb6a4a}
\definecolor{weak_neg}{HTML}{fcae91}

\definecolor{negligible}{HTML}{FFFFCC}

\newcounter{findingcounter} 

\newcommand{\rev}[1]{{#1}}
\newcommand{\revtwo}[1]{\textcolor{black}{#1}}

\newcommand{\finding}[1]{
\refstepcounter{findingcounter}
\begin{center}
\vspace*{-0.3cm}
\setlength{\leftskip}{0pt}
\begin{tikzpicture}
\node [draw=none, fill=gray!20, rounded corners=3pt, inner sep=4pt, minimum height= 20pt] (box) {
\begin{minipage}{0.965\linewidth}
\textbf{Finding {\thefindingcounter}:} {\rev{#1}}
\end{minipage}
};
\draw[line width=2pt,  color=darkgray] (box.north west) -- (box.south west);
\end{tikzpicture}
\vspace*{-0.3cm}
\end{center}
}

\begin{document}

\let\WriteBookmarks\relax
\def\floatpagepagefraction{1}
\def\textpagefraction{.001}

\shorttitle{SE Higher Education in the Netherlands}
\shortauthors{Heeren et al.}

\title [ ] {\Large A Systematic Analysis of Higher Education on Software Engineering\\ in the Netherlands}


\author[ou]{Bastiaan Heeren}[orcid=0000-0001-6647-6130]

\author[uu]{Fabiano Dalpiaz}[orcid=0000-0003-4480-3887]
\ead{f.dalpiaz@uu.nl}

\author[tue]{Mazyar Seraj}[orcid=0000-0002-4108-3186]
\ead{m.seraj@tue.nl}

\author[unifi]{Roberto Verdecchia}[orcid=0000-0001-9206-6637]
\ead{roberto.verdecchia@unifi.it}

\author[ut]{Vadim Zaytsev}[orcid=0000-0001-7764-4224]
\ead{v.zaytsev@utwente.nl}

\affiliation[ou]{%
  organization={Open University},%
  country={The Netherlands}%
}

\affiliation[tue]{%
  organization={Eindhoven University of Technology},%
  country={The Netherlands}%
}

\affiliation[unifi]{%
  organization={University of Florence},%
  country={Italy}%
}

\affiliation[ut]{%
  organization={University of Twente},%
  country={The Netherlands}%
}

\affiliation[uu]{%
  organization={Utrecht University},%
  country={The Netherlands}%
}


\begin{abstract}
~\textbf{Objectives:} 
Software engineering educators strive to continuously improve and refine their courses and programs. Understanding the current state of practice of software engineering higher education can empower educators to critically assess their courses, fine-tune them, and ultimately enhance their educational curricula. In this study, we provide an encompassing analysis of higher education on software engineering by considering the educational offering of the Netherlands.\newline
\textbf{Study methods:}
We adopt a crowdsourced analysis considering 10 Dutch universities and 207 courses. Courses are analysed \textit{via} a set of key knowledge areas adapted from the SWEBOK, which are mapped to courses by educators of their universities. The mapping process is refined \textit{via} homogenisation and internal consistency improvement phases, followed by a data analysis phase.\newline
\textbf{Findings:}
Given its fundamental nature, Construction and Programming is the most covered knowledge area at Bachelor level. Other knowledge areas are equally covered at Bachelor and Master level (e.g., software engineering models), while more advanced ones are almost exclusively provided at Master level (e.g., Maintenance). Three clusters of tightly coupled knowledge areas emerge: (i) requirements, architecture, and design, (ii) testing, verification, and security, and (iii) process-oriented and DevOps topics. Dutch universities cover all knowledge areas uniformly, with minor deviations reflecting institutional research strengths.\newline
\textbf{Conclusions:} 
Our results highlight correlations among key software engineering knowledge areas. We also identify underrepresented areas, such as software economics, which educators may consider including in curricula. We invite researchers to make use of our research method in their own geographical region to globally compare software engineering education programs.
\end{abstract}



\maketitle



\section{Introduction}\label{sec:intro}
Software engineering (SE) education plays an important role in preparing the next generation of software professionals and researchers. Software is at the basis of all kinds of systems performing bank transactions, transportation and logistics tasks, healthcare and insurance activities. Universities want their graduates to possess solid foundational knowledge and practical skills to build and maintain software-intensive systems. Institutions of higher education thus carry a significant responsibility to provide \SE courses that align with the current industry demands, technological trends and research advancements. 

\rev{
However, defining the contents of a SE curriculum and of SE courses entails delicate trade-offs that should account for international guidelines~\cite{ardis2015se,CS2023}, standard bodies of knowledge~\cite{SWEBOKv4}, emerging trends (e.g., sustainability~\cite{chitchyan2025embeddingsustainabilitysoftwareengineering}, artificial intelligence~\cite{hassan2024towards}), academia-industry gaps~\cite{akdur2022}, and the research strengths of the academic institution (especially in post-graduate programs). The premise of our study is that \textit{different academic institutions make different trade-offs and, therefore, we should expect visible differences.}}

This paper presents the findings of a nationwide survey of \SE courses offered at Dutch universities, with the goal of understanding how these courses map to established bodies of knowledge and, by doing so, identifying opportunities for curriculum improvement and course sharing.
This collaborative initiative stems from the VERSEN network --- the Dutch National Association for Software Engineering\footnote{\url{https://www.versen.nl}}. Within VERSEN, the Working Group on Education\footnote{\url{https://www.versen.nl/contents/works/education}} focuses on continuously improving SE education across the Netherlands. 
\rev{We study a single country because research has shown the perils of conducting cross-country comparisons in higher education~\cite{teichler2014opportunities}: while these would provide very valuable knowledge, there are significant contextual factors that characterize specific countries (see \autoref{sec:general}). Although a single-country study is more limited, it can be seen as a multi-case study with a relatively homogeneous sample: in our case, \textit{research universities} in the Netherlands.}

Dutch higher education is governed by \emph{Dutch Higher Education and Research Act}, officially known as the \emph{Wet op het hoger onderwijs en wetenschappelijk onderzoek} (WHW)~\cite{WHW}. The WHW defines the structure of university programmes and diplomas, as well as core regulations on study units and the European Credit Transfer and Accumulation System (ECTS). Under this system, Bachelor of Science (\bachelor, undergraduate) programmes typically consists of 180 ECTS, corresponding to three years of full-time study.
These \bachelor programmes often culminate in a substantial project, carried out either individually or in a small group --- demonstrating students' ability to synthesise \SE concepts and practices. 
Master of Science (\master, graduate) programmes span either one or two years (60 or 120 ECTS, respectively), depending on the university and the specific programme design. \master programmes, besides going deeper into the content, are expected to place greater emphasis on critical thinking, research skills, and the ability to contribute new insights to the discipline~\cite{MeijersCriteria}. Consequently, about one-quarter of an \master student's time is dedicated to an individual final project that focuses on applying advanced topics and tackling research-oriented challenges.


Dutch universities organize the academic year into either two semesters or four quartiles.
Each course confers a certain number of ECTS, and is organised around a set of competencies (generally between 5 and 7.5 ECTS), topics and learning objectives, which are explicitly listed in the course descriptions we will be analysing in the following sections. To help students navigate the landscape of courses offered within their university, \bachelor courses are often grouped into modules~\cite{Visscher2017,Vision2023}, themes, learning lines~\cite{UU-leerlijnen,UvA-leerlijnen}, etc., and \master courses are organised into tracks or specialisations. We elaborate on it more in the next section where we introduce all the universities covered in this survey.

In this research, we classified the content of these courses against the \emph{Software Engineering Body of Knowledge (SWEBOK) v4}, published by the IEEE Computer Society~\cite{SWEBOKv4}. While we acknowledge alternative frameworks such as the ACM/IEEE \emph{Computer Science Curricula} (CS2023)~\cite{CS2023} and other discipline-specific guidelines, SWEBOK provides a widely recognised reference model specific for \SE. It organises \SE knowledge into distinct \emph{knowledge areas} (KAs), offering a systematic way to map, compare, and analyse the coverage of \SE topics across multiple institutions.

\rev{This article makes two contributions:}
\begin{itemize}
    \item \rev{We devise a systematic method for mapping the offering of software engineering in higher education;}
    \item \rev{Via the application of the method to ten universities in the Netherlands, we reveal similarities and differences across knowledge areas and between universities.}
\end{itemize}


\textit{Paper organization}
\autoref{sec:context} provides an overview of higher education in SE in the Netherlands, defining the scope of our analysis. 
\autoref{sec:rgq} presents our research goal and questions, followed by our research method in \autoref{sec:research_method}. 
\autoref{sec:categories} reports our findings per knowledge area ($\mathbf{RQ_1}$), while \autoref{sec:rq2} explores correlations between these areas ($\mathbf{RQ_2}$). 
\autoref{sec:rq3} analyses university educational foci in SE ($\mathbf{RQ_3}$). 
\autoref{sec:rel_work} discusses related work, and \autoref{sec:discussion} includes a broader discussion, threats to validity, and implications for SE education. 
Finally, \autoref{sec:summary} provides a summary and directions for future research.

\section{Context: Higher Education on \SE in the Netherlands}\label{sec:context}
We discuss higher eduction on \SE in the Netherlands to contextualise our study. After some remarks on the general context in \autoref{sec:general}, we provide insights on \bachelor programmes in \autoref{sec:ba} and \master programmes in \autoref{sec:ma}, following by a definition of the scope of our analysis in \autoref{sec:scope}.

\subsection{General Context}
\label{sec:general}
Dutch higher education institutions can be broadly classified into two categories: (i) \textbf{research universities} (\textit{universiteiten}) for research-oriented higher education and (ii) \textbf{universities of applied sciences} (\textit{hogescholen}) for higher professional education~\cite{SK123}. 
The former category has a further sub-class, technical universities (Delft, Eindhoven, Wageningen and Twente), which have a clear focus on technical subjects and on the engineering discipline~\cite{4TU}. 
%
Unlike research universities, universities of applied sciences prioritise skills-based learning, often in collaboration with industrial partners. 
%
\bachelor-level graduates of research universities can freely enrol in many \master-level programmes in their field~\cite[Article 7.30b]{WHW}, while \bachelor-level graduates from universities of applied sciences are often required to follow a short premaster programme (six months or a year)~\cite[Article 7.30e]{WHW} to catch up on theoretical knowledge. 
For computer science, such a premaster typically includes several subjects in mathematics.

The landscape of \SE education in the Netherlands offers many options to study seekers both from the country as well as from abroad. \SE knowledge is offered by most Dutch higher education institutions, primarily within computer science programmes and departments.
\emph{Computer Science} and \emph{Technical Computer Science} (\emph{Informatica} and \emph{Technische Informatica}, respectively) are established terms that attract students since the 1980s, and diplomas of such programmes are still actively sought in after the 2020s. 
These are complemented by programmes like \emph{Business and IT} or \emph{Information Science} (\emph{Informatiekunde}), as well as variations of \emph{Data Science} and/or \emph{Artificial Intelligence}. 
There are then several smaller thematic \bachelor programmes like \emph{Medical Computer Science} and \emph{Computational Social Sciences} (both at the University of Amsterdam); similarly, at the \master level, there are specialised programmes like \emph{Digital Forensics} at the University of Leiden or \emph{Computer Security} at the Vrije Universiteit Amsterdam.
While this reflects the field's diversity and the willingness of higher education to be responsive to emerging technological advancements as well as the market need, in this study \textbf{we focus mostly on the (technical) computer science programmes} where \SE courses play a central role.

\subsection{BSc programmes}
\label{sec:ba}
There are no specific \SE programmes at the \bachelor level. As explained earlier, \SE education is embedded in (technical) computer science programmes. 
The most visible distinction concerns the language of education, which can be English or Dutch. Programmes taught in Dutch are popular at institutions with an established student influx, often from within a particular region of the country, and are motivated by the perceived ease of learning in one's native language~\cite{SoosaiRaj2018} as well as the regional job market needs. Examples are the computer science programmes of Open University, Utrecht University, University of Amsterdam, and Leiden University, having cohort sizes of 148 to 221 students in 2024~\cite{Keuzegids-BSc}.
Programmes in English are motivated by attractiveness for international applicants, but also appealing for Dutch students aiming at international careers. Universities like Radboud University in Nijmegen, University of Groningen, Vrije Universiteit Amsterdam, Eindhoven University of Technology, and University of Twente teach and run their programmes in English. These programmes often see a broader influx, such as Vrije Universiteit Amsterdam, which enrolled 542 students this year~\cite{Keuzegids-BSc}. The smallest programme among these is the one offered in Groningen, with 121 students, the other programmes are at 204--289 students~\cite{Keuzegids-BSc}. The Technical University of Delft recently took the decision to take a hybrid approach, offering parallel influx tracks in Dutch and English, attracting 474 students overall~\cite{Keuzegids-BSc}.
The decision to teach in English often hinges on strategic factors such as long-term growth, and even though the cultural shift associated with English-language instruction and the multiculturalism that brings its own challenges, it does simplify hiring of teaching personnel. 

Some \bachelor programmes limit student intake via the so-called \emph{``Numerus Fixus''}, an admission cap based on an entrance exam result, regulated by a separate law, called \emph{Regeling aanmelding en toelating hoger onderwijs} (Regulations for Registration and Admission to Higher Education)~\cite{RATHO}.
This measure places a legal limit on enrolment numbers, and students are only invited to join the programme in the order of their ranking based on their performance on the entrance exam. Institutions such as University of Groningen, Vrije Universiteit Amsterdam, Eindhoven University of Technology, and Delft University of Technology have adopted this approach to control intake and maintain quality standards in the face of rising demand. 
At the \bachelor level, this is the only admission and selection procedure which allows a university to guarantee a cap on their student influx.

\subsection{MSc programmes}
\label{sec:ma}
\master programmes across the Netherlands range from general computer science curricula to highly specialised offerings. General computer science programmes, offered by most universities, typically provide tracks or specialisations\footnote{Tracks are ministry-regulated, and specialisations can have different definitions per university~\cite{Tracks}.} that are more closely aligned with SE, such as \emph{Software Science} at the University of Groningen, \emph{Software Technology} at the University of Twente and \emph{Software and Analytics} at Eindhoven University of Technology. These tracks or specialisations enable students to focus on core SE competencies within a broader computer science framework, allowing for the integration of SE principles with foundational computer science. Yet, specialised programmes such as the University of Amsterdam's or Open University's \master in \emph{Software Engineering}, aim to provide a deeper, more targeted curriculum that aligns with industry trends in automation, agile practices, and advanced software testing methods.

Several institutions offer \master degrees in \emph{Business and IT} or related somewhat interdisciplinary programmes, bridging technical and business knowledge. These programmes are officially classified and evaluated in a separate educational category (\textit{Information Science}).

Enrolment sizes across Dutch \master programmes vary significantly, reflecting differences in institutional focus, student demographics, and intake flexibility. Delft University of Technology has the largest related \master programme, with 258 new enrolments in 2024, according to Keuzegids data~\cite{Keuzegids-MSc}, while the joint \master programme between the University of Amsterdam (UvA) and Vrije Universiteit Amsterdam (VU) is the second-largest, inviting 151 students on the same year~\cite{Keuzegids-MSc}. Leiden University, which does not have a dedicated SE track, reported 126 new students in its general computer science program, making it the third-largest. Mid-sized programmes, with enrolments ranging between 69 and 112 students, include those at Twente, Eindhoven, Utrecht, Nijmegen, and UvA's specialised \emph{Software Engineering} \master~\cite{Keuzegids-MSc}.

Note that enrolment figures are challenging to define accurately especially in \master programmes, as students often have multiple entry points and pathways. For example, in Utrecht and Twente, new students can start in either September or February. 
Additionally, unlike \bachelor programmes, where students progress together as a cohort, \master students frequently graduate on individual timelines which reflect their personal study experience.

At the \master level, \SE education in the Netherlands follows a remarkably international orientation, with English serving as the dominant language for study materials, which is also in general typical for computer science education. 
This occurs largely due to the recognition that fluency in English, both in technical vocabulary and in day-to-day communication, is an essential skill for future experts to stay competitive on the global SE market. 
The noticeable exceptions are the Open University's two distinct \master programmes in \emph{Computer Science} as well as in \emph{Software Engineering}, which are taught in Dutch, catering specifically to a domestic audience (as well as covering Dutch-speaking regions of neighbouring Belgium) and offering some flexibility for Dutch-speaking professionals.


\subsection{Scope of our analysis}
\label{sec:scope}
Our research focuses on \SE courses offered within computer science and \SE programmes offered by the following research universities in the Netherlands, listed in alphabetical order, with names of cities that host them, in parenthesis, and abbreviations that we will use from now on to save space:

\begin{itemize}
    \itemsep0em 
    \item Delft University of Technology (Delft) --- \TUD
    \item Eindhoven University of Technology (Eindhoven) --- \TUe
    \item Open University (Heerlen) --- \OU
    \item Radboud University (Nijmegen) --- \RU
    \item Rijksuniversiteit Groningen (Groningen) --- \RUG
    \item University of Amsterdam (Amsterdam) --- \UvA
    \item University of Leiden (Leiden) --- \UL
    \item University of Twente (Enschede) --- \UT
    \item Utrecht University (Utrecht) --- \UU
    \item Vrije Universiteit Amsterdam (Amsterdam) --- \VU
\end{itemize}

We therefore exclude Maastricht University, which did not offer a computer science programme at the time this analysis was performed; Erasmus University Rotterdam, which profiles itself in economics and medical sciences and thus does not provide SE education; Tilburg University, which similarly focuses on humanities and law; and Wageningen University, which focuses on agricultural and environmental research and education. We also exclude all universities of applied science (at least 48) not only to make the work feasible, but also to avoid the discussion on the level of the learning objectives, since they have a more applied and market-driven nature.












\section{Research Goal and Questions}
\label{sec:rgq}

In order to define our goal and research questions, we follow the Goal-Question-Metric approach first suggested by Basili~et~al.~\cite{caldiera1994goal}. Our overarching research goal is defined as follows:

\begin{quote}
\textit{Analyse} the SE educational landscape\\
\textit{For the purpose of} understanding the trends\\
\textit{With respect to} course content\\
\textit{From the viewpoint of} SE educators\\
\textit{In the context of} the Netherlands.
\end{quote}

As outlined in our goal, the objective of this research is to gain a systematic understanding of the current higher education on SE in the Netherlands. As further detailed in Section~\ref{sec:research_process}, such goal is achieved \textit{via} a systematic analysis of course content crowd-sourced across educators of 10 Dutch universities providing SE academic instruction.

Following the GQM approach, we derive the following research questions (RQ), that we need to answer in order to achieve our research goal.

\begin{quote}
$RQ_1$: \textit{What are the topics studied in \SE Courses?}
\end{quote}

We aim to gain insights into the content of the \SE courses provided by the considered Dutch universities. As further specified in Section~\ref{sec:research_process}, to answer such RQ, we focus on the analysis of key \textit{Knowledge Areas} (KAs), initially taken from the SWEBOK, and further refined via an iterative process.
The data collected for $RQ_1$ lays the groundwork to answer our subsequent RQs.

\begin{quote}
$RQ_2$: \textit{What are the (co-)occurrences among studied topics and the educational context?}
\end{quote}

We investigate the potential inter-dependencies between SE topics, not only by considering inter-KA occurrences, but also by taking into account the frequency with which the topics appear among graduate and undergraduate courses. In addition, for $RQ_2$, we also study the role that SE-specific programmes and tracks play in the presence of KA in courses.

\begin{quote}
$RQ_3$: \textit{Do universities differ in terms of education foci?}
\end{quote}

We look for differences in terms of SE educational content across the considered universities\rev{, which we would expect because of the different trade-offs universities make when defining their programs and courses}. By answering $RQ_3$, we can assess the extent to which educational programmes providing SE education display heterogeneity and similarity across their formative offer.

\section{Research Method \rev{and Process}}
\label{sec:research_method}
\rev{We first discuss the broader research methodology aspects of our study in \autoref{sec:meth_emb}, and then detail the research process we followed in \autoref{sec:research_process}.}

\subsection{\rev{Methodological Embedding}}
\label{sec:meth_emb}

\rev{Our research method is a combination of: (i) \textit{cross-case analysis}~\cite{borman2012cross}, as we are comparing multiple universities; and (ii) \textit{deductive qualitative analysis}~\cite{gilgun2019deductive}, as we analyze collected data in a predominantly qualitative manner according to an existing theory/framework: the knowledge areas in the SWEBOK. Although these techniques are often used within grounded theory~\cite{glaser1998grounded}, we cannot argue we are conducting that research method because we do not construct a theory as part of our analysis.}

\rev{Our study can be seen under the lens of \textit{comparative higher education research}~\cite{teichler2014opportunities}.
The field, as observed by Teichler~\cite{teichler2014opportunities}, is characterized by an \textit{inherent tension} between obtaining a rich contextual understanding (best addressed in a single institution or country) and performing a comparison at the international level.  Teichler explains how national studies are often seen as ``over-descriptive'', while international studies tend to be shallow because of limited contextual knowledge, they involve a single researcher per country, they suffer from a coincidental choice of countries,  often a single foreign (conveniently selected) country. Indeed, Kosm{\"u}tzky and Kr{\"u}cken~\cite{kosmutzky2014growth} found that 67\% of the international studies compare only two countries. The debate on how to best tackle the tension between context and comparability is still open~\cite{kosmutzky2020between}.}

\rev{
This study is a first step toward \textit{theory building}: our analysis offers a replicable process (see \autoref{sec:research_process}) and the key results that we report in this paper are highlighted as a set of \textit{findings}. These findings derive from a national context that involves every research university in the Netherlands with a focus on SE education. We involve multiple researchers from one country, based on their profound knowledge of the regional context (\autoref{sec:context}), and we crowdsource data collection to the universities themselves.
We acknowledge that \textit{theory testing} is an important future direction: to what extent are our findings specific to the context of the Netherlands, to the time snapshot of our analysis, and to other factors? We encourage researchers from other countries to replicate the study either at the national or international level; in the latter case, particular attention must be placed on achieving rigor in the selection of countries and institutions~\cite{kosmutzky2016precision}.}

\subsection{\rev{Research Process}}
\label{sec:research_process}
The research process we followed, depicted in Figure~\ref{fig:research_process}, is composed by the following steps:

\begin{figure*}[pos=h]
    \centering
    \includegraphics[width=1\linewidth]{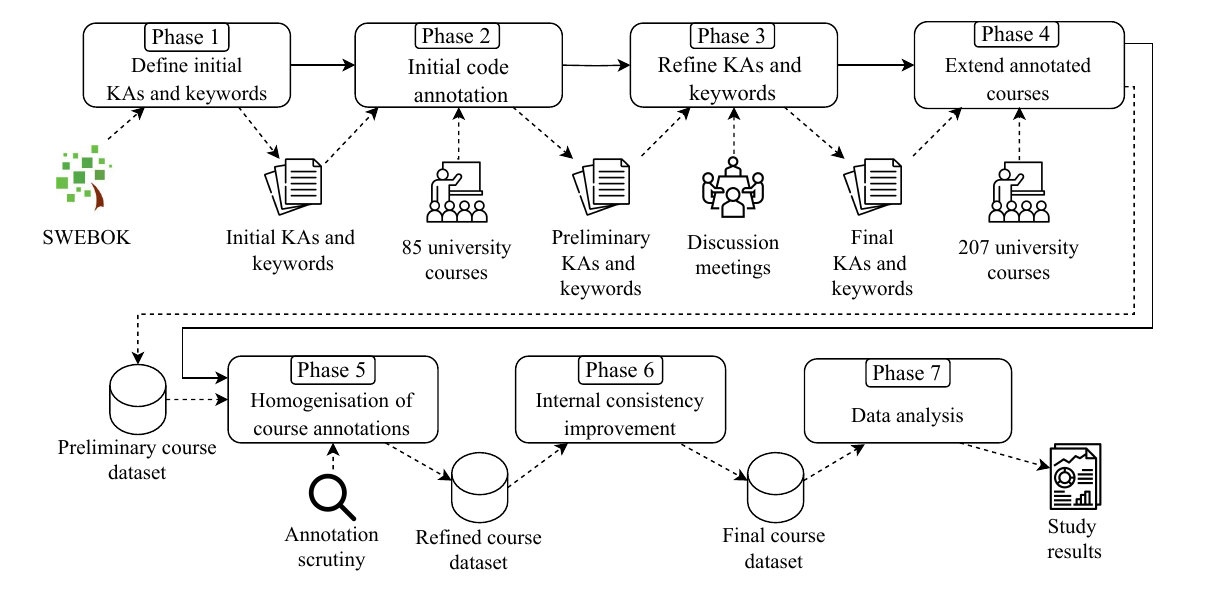}
    \caption{Research process overview}
    \label{fig:research_process}
\end{figure*}
\begin{enumerate}
    \item \textit{Define initial KAs and keywords.} Two of the five authors of this paper have initiated the process by identifying a reference framework for categorising the courses according to knowledge areas of SWEBOK. At the time of our data analysis, SWEBOK was in a phase of transition from version 3 to version 4. As such, these authors considered the structure and a draft of the SWEBOK V4, which provided a more up-to-date view of the discipline\footnote{\rev{For examples, SWEBOK V4 decouples software architecture from software design, a distinction that has been argued for since the 1990s~\cite{perry1992foundations}.}} including 18 knowledge areas (KAs) than SWEBOK v3 which dates back to 2014. 
    Out of these 18, the authors selected the 15 key KAs: \req, \arch, \des, \ka{Construction}, \test, \ops, \maint, \conf, \semgmt, \seproc, \ka{Models and Methods}, \ka{Quality}, \ka{Security}, \ka{Professional Practice}, and \economics. We excluded the KAs that represent foundational knowledge: Computing Foundations, Mathematical Foundations, and Engineering Foundations. For each of the 15 KAs, the authors have read the draft chapter of SWEBOK V4 and have identified a set of keywords that are likely to characterise key topics within the KA. For example, for the KA \req, we identified \req, `Elicitation', `Specification', `Analysis', `System objectives', `Use cases', `Non-functional requirements', etc.
    \item \textit{Annotate initial set of courses.} Three authors (the two involved in step (1) and an additional one) collected and annotated --- via a spreadsheet --- collections of SE-specific courses from four universities: \OU, \TUe, \UU, and \VU. These were gathered either directly by the authors being an employee of a university, or through an examination of the course catalogues with the help of a colleague working at that institution. The annotation process consisted of: (a) reading thoroughly the course description; (a) ticking the KAs that the course contributes to, based on the presence of the keywords for that KA or related terms, and (c) taking notes about the course or the annotation process, when necessary. We did not employ multiple annotators for these 85 courses; however, the three authors checked each others' annotations and had multiple rounds in which they discussed discrepancies, leading to the next step.
    \item \textit{Refine KAs and keywords.} Based on the experience gained from the annotation process and from the discussion meetings, the three authors involved in step (2) worked at a revised set of KAs and keywords. A few major changes were applied:
    \begin{itemize}
        \item Given the high number of courses related to programming, we have extended the \ka{Construction} KA to \cnp.
        \item Since the methodological component of the KA \ka{Models \& Methods} overlaps with \semgmt and \seproc, we restricted the KA to modelling and model-driven engineering, and we use the term \modeling.
        \item We removed two KAs: \ka{Software quality}, as it can be argued that most KAs contribute to software quality; and \ka{Professional practice}, as this was found too seldom in the examined courses.
        \item We introduced two additional KAs that represent an important aspect of SE education in the Netherlands and that are not listed as KAs by the SWEBOK: \verif and \ka{Programming language design} (briefly: \pldesign). 
        \item We introduced an additional field that denotes the type of course: (a) \prog denotes a course teaching a programming paradigm or language; (b) \pldesign groups the courses that focus on the design of programming languages; (c) \seminar indicates research courses; (d) \senum refers to introductory courses on SE; (e) \proj indicates courses where the learning is mainly achieved by conducting a project~\cite{Gary2015} for a real or simulated client.
        \item We removed six courses (from the 85 we analysed) that were deemed irrelevant.
    \end{itemize}
  
    \autoref{tab:ann-scheme} provides an overview of the final annotation scheme; the top part shows the KAs, while the bottom part focuses on the clusters of courses.

\begin{table*}[pos=h]
\centering
\caption{Overview of the final annotation scheme. The upper part shows KAs, the lower part focuses on course clusters.}\label{tab:ann-scheme}
\footnotesize
\begin{tabular}{lp{12cm}}
\toprule
Knowledge Area                          & Keywords/topics    \\
\midrule
Requirements                      & requirements, elicitation,   specification, analysis, system objectives, goals, user stories, use cases,   non-functional requirements                                                                                                                         \\
Architecture                      & software architecture, ADL, 4+1   model, concerns, architectural styles, software product lines, architectural   definition, architectural patterns, quality attributes, architectural tactics                                                                 \\
Design                            & UML, class diagrams,   object-oriented design, software design, design patterns, refactoring,   unified process, domain modelling                                                                                                                               \\
Construction   \& Programming & programming (imperative,   object-oriented, functional, web, game, model-driven), C\#, Java, Haskell,   Scale, Python, HTML, PHP, Javascript, app development, business rules, code   documentation, IDE                                                       \\
Testing                           & testing (unit, integration,   system, acceptance), property-based testing, record-and-replay, mutation   testing, partition testing, white box, black box, test cases, test set,   coverage criteria, testing model, software quality, scriptless, model-based \\
SE Operations                        & continuous delivery, continuous   integration, deployment, integration, packaging, monitoring, release   engineering                                                                                                                                           \\
Maintenance                       & software evolution,   maintenance, change, product quality, software quality management, impact   analysis, technical debt, software re-engineering                                                                                                            \\
Configuration management                & version control, git, svn,   configuration, dependencies, source code management system                                                                                                                                                                        \\
SE management                     & SDLC, project planning, cost estimation, effort estimation, requirements change, schedule, SE measurement                                                                                                                                                    \\
SE process                        & (product) life cycle, methods,   waterfall, iterative, agile, spiral, V-model, Scrum, software process   assessment and improvement, minimum viable product, proof of concept                                                                                  \\
SE models                        & domain modelling, UML,   variability modelling, OCL, business modelling, metamodelling, process modelling,   BPMN, model-driven development                                                                                                                         \\
Software Security                       & security vulnerabilities,   security by design, sandboxing, common vulnerabilities, security risk   analysis, ethical aspects of security, penetration testing, security patterns                                                                              \\
SE economics                      & finance, accounting,   controlling, cash flow, valuation, product vs. project vs. system, costing,   estimation, prioritisation                                                                                                                                \\
Verification                  & verification, LTL, CTL, model   checking, symbolic execution, theorem proving, formal specification, static   analysis, program analysis                                                                                                                       
\\
Progr.\   Language design        & program syntax, semantics,   compilers, types, type systems            
\\\midrule
Course cluster & Description
\\\midrule
SE 101                           & A basic first course on software engineering             \\
Project                           & A project-based course where   the students learn, either solo or in teams, how to apply SE in a practical setting. Often a capstone project.    \\
Seminar & A research-oriented course on advanced SE topics.\\
Programming & A course that teaches a programming paradigm or language. \\
PL Design & A course on the design of programming languages.\\
\bottomrule
\end{tabular}
\end{table*}

    \item \textit{Extend the set of annotated courses.} The revised annotation scheme from step (3) was then used in a second annotation round. The three authors involved in step (2) have re-analysed the courses from the four universities, and at the same time, selected researchers/educators from six additional universities (\RU, \RUG, \TUD, \UvA, \UL, \UT) were invited and asked to (a) provide a list of SE courses taught at their institution, and (b) annotate those using the annotation scheme. This led to an additional set of 207 courses with annotations. \rev{These additional colleagues were selected based on convenience sampling: we invited contacts we had met at events on SE education and with expertise beyond teaching individual courses.}
    \item \textit{Homogenisation of course annotations.} The new set of 207 courses was analysed by the same three authors in order to identify discrepancies from the annotation that was performed on the 79 courses (which were left over from 85 when six were removed). We observed that many courses were prerequisites for SE courses but could not be mapped to the KAs we included; this led to removing 103 courses, thereby retaining 128 courses for the additional six universities. Summing up to the annotated courses from the first four universities, this results in 207 courses that are included in this analysis. During this process, the three authors also checked the ticked KAs against the course descriptions, and made several adjustments whenever they saw a discrepancy from the annotation scheme. We have found, indeed, that several KAs had been ticked even when that KA was not a primary focus for that course. In total, this resulted in 167 removed ticks and 39 added ones. Given the 319 agreements to tick and the 2,580 agreements to not tick the KA, this results in a Cohen's kappa of 0.718, which represents substantial agreement. A detailed analysis of the inter-rater agreement per category can be found in the online appendix\footnote{Our online appendix includes our coded data as well as other scripts used for our analysis: \rev{\url{https://doi.org/10.5281/zenodo.21276768}}}. \rev{We sent the list of retained courses and our revised annotations to the colleagues who provided the initial lists, giving them the opportunity to challenge our decisions; we have not received objections, only a couple requests for clarifications.} 
    \item \textit{Improve internal consistency.} In order to improve the internal consistency of the tagging, the three authors subdivided the KAs and they checked all 207 courses, with the aim of determining the consistency of the annotations for each KA. \rev{The goal was that of ensuring that a single annotator would make a `vertical' analysis of all courses against one individual KA, looking for omitted or erroneous ticks.} This led only to three adjustments: one tick was added, two were removed. We also identified \textit{signature courses}; we labeled a course as \textit{signature for a KA} if the course revolves almost exclusively around that KA (e.g., a `Requirements Engineering' course would be a signature course for the KA \req).
    \item \textit{Analyse the data.} The resulting spreadsheet was used at the basis of the follow-up analysis, which revolved around the research questions described in \autoref{sec:rgq}. 
\end{enumerate}

\section{Findings per Knowledge Area ($RQ_1$)}
\label{sec:categories}
\begin{figure}[pos=h]
    \centering
    \includegraphics[width=1\linewidth]{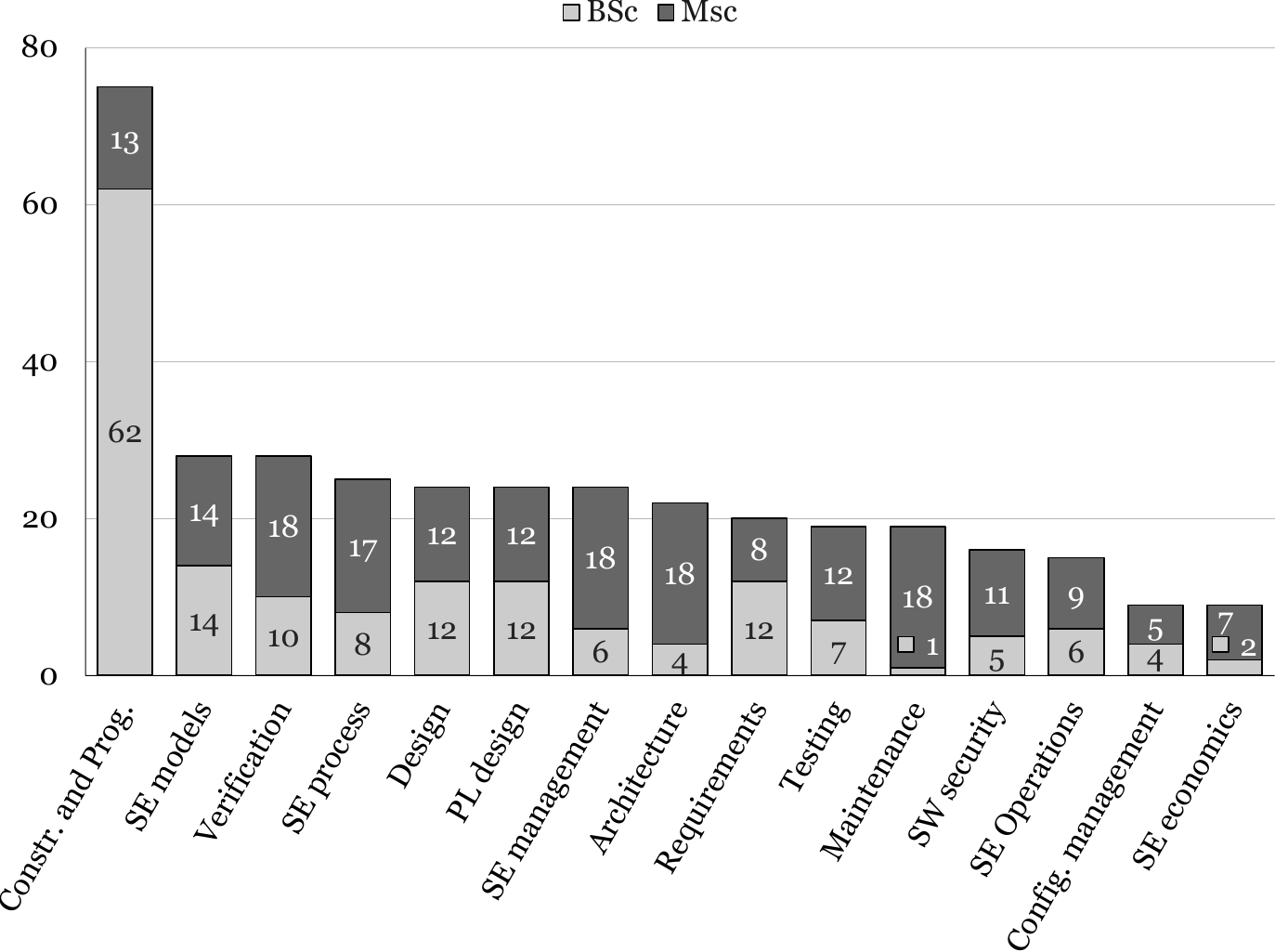}
    \caption{Recurrence of topics (KA) treated in the courses.}
    \label{fig:ka_occurrence}
\end{figure}

We present findings per each knowledge area we analysed, discussing the number of courses per KA, their distribution across \bachelor and \master programs and tracks, course type, and an analysis of the course topics through an analysis of the most frequently occurring topics. For this last activity, we used the the \texttt{wordcloud} package\footnote{\url{https://pypi.org/project/wordcloud/}} to generate the top 20 terms in the KA, also including two-word compounds, after removing stopwords, and applying lemmatisation. For the sake of brevity, we present 
A summary of the recurrence of the KAs across courses is shown in \autoref{fig:ka_occurrence}.

\subsection{Requirements}

\begin{table*}[pos=h]
\scriptsize
\caption{\rev{Top twenty common terms for the KAs \req, \arch, \des, \test, and \maint with relative frequency compared to the most frequently occurring term in that KA.}}
\label{tab:top-words-1}
\setlength{\tabcolsep}{2pt}
\begin{tabular}{llllllllll}
\toprule
\req & Freq. & \arch & Freq.          &  \des & Freq. & \test & Freq. & \maint & Freq. \\
\midrule
requirement              & 1.000 & architecture          & 1.000 & design               & 1.000 & testing            & 1.000 & software               & 1.000  \\
technique                & 0.708 & software architecture & 0.791 & software             & 0.958 & test               & 0.544 & software evolution     & 0.548  \\
specification            & 0.583 & software              & 0.674 & model                & 0.583 & software           & 0.533 & software system        & 0.452  \\
software                 & 0.542 & system                & 0.535 & system               & 0.552 & system             & 0.478 & quality                & 0.355  \\
system                   & 0.458 & design                & 0.442 & architecture         & 0.396 & technique          & 0.444 & metric                 & 0.290  \\
user                     & 0.375 & requirement           & 0.279 & analysis             & 0.344 & model              & 0.389 & technique              & 0.258  \\
need                     & 0.375 & concern               & 0.209 & engineering          & 0.333 & software testing   & 0.233 & code                   & 0.258  \\
elicitation              & 0.333 & stakeholder           & 0.209 & requirement          & 0.313 & code               & 0.189 & tool                   & 0.226  \\
requirement engineering  & 0.333 & quality               & 0.209 & use                  & 0.292 & quality            & 0.189 & engineering            & 0.194  \\
diagram                  & 0.292 & project               & 0.186 & tool                 & 0.281 & risk               & 0.178 & maintenance            & 0.194  \\
process                  & 0.250 & architect             & 0.186 & problem              & 0.260 & program            & 0.167 & requirement            & 0.194  \\
practice                 & 0.250 & structure             & 0.163 & application          & 0.260 & use                & 0.156 & language               & 0.194  \\
development              & 0.208 & model                 & 0.163 & software system      & 0.260 & design             & 0.144 & method tool            & 0.194  \\
use                      & 0.208 & component             & 0.140 & design pattern       & 0.250 & process            & 0.133 & evolution              & 0.161  \\
formal                   & 0.208 & practical             & 0.140 & method               & 0.229 & tool               & 0.122 & change                 & 0.161  \\
method                   & 0.167 & style                 & 0.140 & principle            & 0.229 & specification      & 0.111 & development            & 0.161  \\
informal                 & 0.167 & decision              & 0.140 & project              & 0.219 & analysis           & 0.111 & program comprehension  & 0.161  \\
project                  & 0.167 & concept               & 0.140 & software development & 0.219 & state              & 0.100 & evolution discus       & 0.161  \\
activity                 & 0.125 & development           & 0.140 & programming          & 0.198 & method             & 0.100 & step                   & 0.129  \\
stakeholder              & 0.125 & analysis              & 0.140 & quality              & 0.198 & modelbased testing & 0.100 & analysis               & 0.129   \\
\bottomrule
\end{tabular}
\end{table*}

Looking at the coverage of the \req KA, we see that nine of the ten universities considered cover this knowledge area in their courses, offering a total of 20 courses. Twelve of these courses are offered to \bachelor students. Eight courses are taught in SE-specific programs or tracks. Three of them are at \UvA as part of their \master in \emph{Software Engineering}. All 20 courses are regular ones, with no specific course type such as `Seminar' or the like.


Five universities (\RU, \UL, \UU, \UvA, and \VU) have signature courses that include the term \ReqEng\ (RE) in the course name. Among the other universities, \TUe has a highly related course called \SSpe. Out of these six courses, four are taught to undergraduate students, and two (\UU, \UvA) are offered to graduate students.

The first column of \autoref{tab:top-words-1} presents the most frequent terms and shows that, in addition to the obvious term `requirement', highly prevalent topics refer to the fact that `techniques' are being taught (as opposed, e.g., to algorithms or theories), the `specification' phase, the focus on `software' as well as on `systems'. The words `user' and `need' highlight the importance of considering the perspective of the actual users of the system, and the fact that these have needs, rather than ready-made requirements. Other interesting terms are `elicitation' and `diagram', classical phases and techniques that are taught in these courses.

\subsection{Architecture}
\label{sec:ka-architecture}
There are 22 courses offered by the ten examined universities; just like for requirements, all universities offer at least one course that covers this KA. However, only four courses are at the \bachelor level, indicating that software architecture is deemed as an advanced topic in the context of SE education. 15 courses are part of an SE-specific track or program.

There are six signature courses that focus solely on software architecture (\OU, \RUG $\times2$, \TUD, \UT, \UU). They are \textit{all} at the \master level, also indicating the advanced nature of the topic in the analysed curricula. Interestingly, \RUG offers two courses on software architecture; one of which, denoted as \textit{advanced}, has a clear focus on preparing professionals, also thanks to the emphasis on a project. Two courses are offered as research seminars: \BIS\ (\UL) and \FAS\ (\VU).


Some of the analysed courses take a perspective that goes beyond pure software, and they enter the field of information systems. While software architecture is a core topic in SE, architectural design is also a prominent activity for other design artifacts. This is visible, for example, in the following courses: \SBIS\ (\UL), \ICTA\ (\UL), \AIS\ (\UT), and \DA\ (\VU).

The second column of \autoref{tab:top-words-1} highlights that, in addition to the obvious keywords `architecture', `software architecture', `software', `system', and the aforementioned `design', the following most prominent topic is that of `requirement'. This witnesses the relationship between the two knowledge areas~\cite{nuseibeh2001weaving}. Other frequent terms are `concern', `stakeholder', `quality', `structure'. These clearly denote key terms in the field such as stakeholders' concerns, architecture as a structure, and quality aspects.

\finding{While equally covered overall, the \req KA is covered both at the \bachelor and \master level, while the \arch KA is mainly included in \master level courses, indicating that this KA is probably seen as a more advanced topic by curriculum designers. This is in line with observations that teaching this KA is difficult~\cite{galster2016makes}.}

\subsection{Design}
There is a total of 24 courses that are offered by nine of the ten universities in our list. Half of these are at the \bachelor level, half at the \master level. Seven universities have at least one \bachelor course, seven have at least one \master course. Only four universities (\OU, \RUG, \UT, \VU) have dedicated courses on this topic, which links early phases like requirements engineering and software architecture with programming, at both levels. Interestingly, three courses are in the cluster `Programming', indicating that design and programming are sometimes taught together.

Defining signature courses for this category is not simple, as \des is often taught in conjunction with other courses. Six highly specific courses can be identified, out of which three have a clear focus on this category: \SD\ (\TUe and \VU) and \PA\ and \SD\ (\RUG). Other three courses focus on advanced design aspects: \DC\ (\OU), \SDE\ (\TUe), and \SOD\ (\VU). Half of these courses are part of SE programs or tracks.


When we explore the most frequent terms in the descriptions of the courses in this category (\autoref{tab:top-words-1}, third column), we see that the term `software' stands out much more prominently than for the previously analysed KAs \req and \arch. Other major keywords are obviously `design' itself, `model' (linking to the \modeling KA), `system' (indicating a broader perspective than software only), 
`architecture', `analysis', `engineering', and `requirement'. 

\subsection{Construction \& Programming}
This KA is the most frequent: a total of 75 courses across all ten analysed universities cover this KA. The undergraduate courses are by far more popular: 62 of the 75 courses are at the \bachelor level, only 13 are at the \master level. This is in line with the key idea that learning programming is a core competence of any computer science student and, as such, it has to be covered at the \bachelor level. This is further confirmed by the fact that only 12 courses are part of a SE-specific track or program.
Within the 75 courses, 55 are also in the cluster `programming', denoting a course with the main goal of learning a programming paradigm or language. 

By analysing the 13 \master level courses, we observe that these cover advanced programming paradigms, including parallel programming (\RUG and \VU), web and cloud computing (\RUG), cryptography and security (\TUD, \UvA, \UL, \UT), and functional programming (\UU, \RU, \UvA).

\finding{\revtwo{At the basis of SE education lies \cnp, which appears in most \bachelor SE degrees (63/75 occurrences) in courses that focus exclusively on programming (55/75). As a foundational aspect, programming syntax is the first stepping stone towards an educational path into SE knowledge.}}

Given the foundational role of \cnp for computer science and SE, we propose that further studies could be conducted with a sole focus on analysing how programming is being taught across universities, in line with previous studies on the choice of the programming language for introductory courses~\cite{murphy2016analysis,simon2018}.



\subsection{Testing}
\label{sec:ka-testing}
In total, 19 out of the 207 total courses considered present a focus on the \test KA. In terms of universities, \test courses are offered in the vast majority of the universities included in this study (nine out of ten). Similar to the \arch KA (\autoref{sec:ka-architecture}), most courses with an emphasis on software testing are provided at the \master level (12 out of 19), while only a third at \bachelor level. This could be attributed to the impossibility to dig deep into testing concepts while conducting a general purpose \bachelor programming course, leading to the higher presence of the KA in more advanced courses, which necessarily need to be provided at the \master level.

Many courses (11/19) are part of a SE-specific track or program, indicating that this KA is highly specific to \SE. Most courses are regular ones, with three exceptions: one seminar (\SSSec\ at \UL), one in the programming cluster (\CE\ at \UvA), and one project (\PST\ at \VU). The latter two highlight that some educators emphasise how the value of testing is best understood by practising skills.

\finding{Like \arch, \test is also taught mostly at the \master level, and mainly in SE-specific tracks and programs, signalling a highly specific KA for SE. This aligns with a previous study~\cite{tramontana2024state} that shows how software testing courses in Belgium, Italy, Portugal and Spain are mostly taught at the \master level, with a few exceptions in the final \bachelor years.}

By considering signature courses, we notice these to be the majority (12 out of 19). Most are provided at the \master level (9 out of 12), further corroborating that courses putting high emphasis on testing require a more advanced understanding of SE practices, and hence are considered mostly for \master academic curricula.


The most frequent keywords in the course descriptions (fourth column in \autoref{tab:top-words-1}) are the obvious `testing', `test', `software', and `system'. Other frequent keywords include `technique', as testing is often taught as a set of techniques that can be used; `model', probably referring to model-based testing; `code' and `program', indicating against which artifact tests are run; and `quality' and `risk', as testing is proposed as a component of quality assurance~\cite{naik2011software} that can reduce the risk of failure.




\subsection{SE Operations}
The \ops KA is covered in 15 out of 207 total courses, and is considered to different extents in the educational programmes of all ten universities considered for this study. The university that provides the higher number of courses with an \ops component is \UL with two \bachelor and two \master courses considering the topic. In terms of \bachelor and \master courses, we note that the KA is more recurrent at \master level (9 out of 15 total courses with a component focusing on \ops), while being also quite recurrent at \bachelor level (6 out of 15).  Circa half of the courses (7) are part of SE tracks or programmes. Notably, \ops are taught in a variety of methods, including an \senum course, a project, and two seminars.

The courses in the \ops KA are seldom considered as signature courses (2 out of 15), exceptions being \SC\ at \VU and \DevCS\ at \UvA. 
Given the low number of signature courses, we do not create a word cloud and we do not analyse the recurring keywords, as they would be influenced too heavily by a single course.




\subsection{Maintenance}
Out of the 207 courses considered in this study 19 focus on the \maint KA. In terms of universities, courses with a maintenance components are provided in seven out of ten universities, with \TUD resulting the university with more courses focusing on the topic (4 \maint courses). 
Almost all courses are provided at the \master level, with only one course, namely \SEM\ at \TUD, considering the \maint KA at the \bachelor level. 

The signature courses in this knowledge area are tightly coupled with the software evolution topic, and are provided at five different universities: \RUG, \TUe, \UL, \UT, and \UvA.


Given the specialised topic, which could be considered as rather specific to SE, it comes at no surprise that the majority of \maint courses are provided within SE-specific programmes and tracks (15/19).

A minority of courses utilises seminars as learning objectives evaluation (6/19), while only one course, namely the already mentioned \SEM\ at \TUD, has a maintenance component as part of an introductory course on SE.

When we examine the most common words in the signature courses (\autoref{tab:top-words-1}, fifth column), we observe that `software', `software evolution' and `software system' are by far the most significant words, followed by `quality' (maintenance is advocated as a key determinant of software quality), and `metric' (used to measure quality). The following keywords refer to the use of `techniques', the relationship to `code', and the existence of `tools' to perform maintenance.


\subsection{Configuration management}
From our analysis, \conf results overall an infrequent KA, as it is covered in only 9 out of the 207 courses. It is considered in six universities, with \RUG being the only one reporting two different courses focusing on such KA. 

A balance exists between \bachelor (4/9) and \master courses (5/9). Given the specificity of this KA, the inexistence of signature courses is not surprising. Given the overlap with other KAs (see \autoref{sec:rq2}), we do not analyze the most frequent terms.

One of the courses is a seminar, two courses are part of the `programming' cluster, and over the half of the courses (5/9) are taught in SE-specific tracks or programs.


\subsection{SE management}
Unline the previous KA, \semgmt emerges as a rather popular KA, covered in 24 courses. All but one university consider such topic, with \RU, \RUG, and \UL having four courses each. 
Only a fraction of \bachelor courses have this KA (6/24). From an inspection of the \bachelor courses, they all seem to provide a broad overview of \SE\ practices, such as \SEM\ at \TUD and \SP\ at \UvA.

Five signature courses on \semgmt could be identified: \SAM\ at \VU, \SDM\ at \RU, \SPM\ at \TUe, \SM\ at \UT, and \PM\ at \UL (the only signature course at the \bachelor level).

Eleven courses are provided in the context of SE-specific tracks and programs, showcasing the considerable specialisation on SE knowledge the KA implies.

While the majority of the courses considering the \semgmt KA relies on written exams, we also note a slight heterogeneity of learning objectives evaluation methods, with some courses utilising seminars (4/23), projects (2/23), or being part of SE introductory courses (1/23).


\begin{table*}[pos=h]
\scriptsize
\caption{\rev{Top twenty common terms for the KAs \semgmt, \modeling, \security, \economics, and \verif with relative frequency compared to the most frequently occurring term in that KA.}}
\label{tab:top-words-2}
\setlength{\tabcolsep}{2pt}
\begin{tabular}{llllllllll}
\toprule
\semgmt & Freq. & \modeling & Freq.          &  \security & Freq. & \economics & Freq. & \verif & Freq. \\
\midrule
software             & 1.000   & system               & 1.000 & security             & 1.000 & software           & 1.000 & program              & 1.000 \\
project              & 0.800   & model                & 0.786 & software             & 0.520 & development        & 0.431 & system               & 0.500 \\
management           & 0.500   & software             & 0.595 & system               & 0.280 & product            & 0.379 & technique            & 0.424 \\
development          & 0.400   & engineering          & 0.381 & analysis             & 0.260 & management         & 0.362 & software             & 0.390 \\
model                & 0.367   & development          & 0.333 & vulnerability        & 0.240 & software product   & 0.310 & security             & 0.339 \\
process              & 0.300   & tool                 & 0.333 & secure               & 0.240 & software ecosystem & 0.310 & model checking       & 0.314 \\
quality              & 0.200   & design               & 0.333 & computer             & 0.220 & business           & 0.276 & use                  & 0.288 \\
measurement          & 0.200   & information          & 0.310 & software development & 0.200 & model              & 0.224 & testing              & 0.280 \\
issue                & 0.167   & modeldriven          & 0.286 & block                & 0.180 & process            & 0.207 & specification        & 0.280 \\
team                 & 0.167   & language             & 0.286 & programming language & 0.160 & technique          & 0.190 & type                 & 0.280 \\
risk                 & 0.167   & transformation       & 0.238 & level                & 0.140 & software vendor    & 0.190 & tool                 & 0.271 \\
agile                & 0.133   & cps                  & 0.214 & technique            & 0.140 & project            & 0.172 & model                & 0.271 \\
system               & 0.133   & business rule        & 0.214 & principle            & 0.120 & technology         & 0.172 & programming language & 0.263 \\
manager              & 0.133   & modeling             & 0.190 & static               & 0.120 & market             & 0.155 & verification         & 0.220 \\
product              & 0.133   & pattern              & 0.190 & code                 & 0.120 & sustainability     & 0.155 & property             & 0.212 \\
metric               & 0.133   & complex              & 0.190 & problem              & 0.120 & business model     & 0.155 & formal               & 0.212 \\
standard             & 0.133   & problem              & 0.190 & application          & 0.120 & method             & 0.138 & proof                & 0.203 \\
insight              & 0.133   & cyberphysical system & 0.190 & attack               & 0.120 & topic              & 0.138 & correct              & 0.195 \\
cost                 & 0.133   & application          & 0.167 & issue                & 0.100 & measurement        & 0.121 & logic                & 0.195 \\
effort estimation    & 0.133   & specification        & 0.143 & type                 & 0.100 & industry           & 0.121 & prove                & 0.195  \\
\bottomrule
\end{tabular}
\end{table*}

When we analyse the most frequent terms (\autoref{tab:top-words-2}, first column), in addition to the obvious terms `software' and `management', the keyword `process' stands out to indicate the link with managing projects. These are followed by `development' (this is the main `process' being managed), and `model' (referring to maturity and measurement models).


\subsection{SE process}
The KA \seproc is quite popular among SE courses, with 25 out of 207 courses focusing on such KA, and all universities providing at least two courses on it. 

The KA is most treated as part of \master curricula (17/25), while only a single \master course, namely the \textit{Software Process} provided at \UvA, is a signature course on the topic. 

Slightly more than half of the times (15/25), the KA is considered as part of a SE-specific track or program, showing how the topic is important also for students who do not explicitly study SE.

\finding{\semgmt and \seproc are well-covered KAs (24 and 25 courses, respectively): \semgmt is offered in all but one university, \seproc by all universities. This indicates that solid attention is paid to explaining the non-technical aspects of SE. We observe that this is supported by the fact that most popular SE textbooks~\cite{PressmanMaxim2019,sommerville2015software} include ample coverage of these topics.}

\subsection{SE models}
As explained in \autoref{sec:rgq}, the KA \modeling is interpreted here in terms of the use of modelling languages and notations within SE courses.

This is a popular KA, with 28 identified courses, evenly split between \bachelor and \master. Nine of the ten analysed universities offer at least one course in this category; all of them at the \master level, seven of them also through \bachelor courses. 

We could identify five signature courses; in addition to two having obvious names: \MDD\ (\bachelor, \OU); \MDE\ (\master, \UT), we could also find one \bachelor course on \SD\ (\VU), as well as \master courses called \SDE\ (\TUe) and \MDCPS\ (\UvA). All of these signature courses also tick the KA \des, showing the tight interplay between the KAs. 

Among these signature courses, the \TUe course that emphasises modelling in the context of systems engineering and the \UvA course on cyber-physical systems, both emphasising how models are an important abstraction not only for software, but for systems in general. 

\finding{\modeling seems to be a highly represented KA compared to others. This could be explained by the strong focus on model-based SE in the Netherlands, evidenced, e.g., by the involvement of Dutch scholars in the definition of a body of knowledge for model-based SE~\cite{burgueno2019contents}.}


This observation is confirmed by the most occurring terms in the signature courses (\autoref{tab:top-words-2}, second column), where `system' appears as the most frequent term. The second most frequent term is obviously `model' , followed by `software' and `engineering', `development',  `tool', all referring to aspects of modelling paradigm (model-driven engineering (MDE) and development (MDD)). Then, keywords like `design' and `information' follow, denoting the link with design activities and the necessity of modelling information. The next three terms `model-driven', `language', and `transformation' are clearly referring to how languages and transformations are at the heart of MDD and MDE.


\subsection{Software Security}
\label{sec:ka-security}
The KA \security includes a total of 16 courses that are taught by nine of the ten covered universities. The topic seems to be more popular at the \master level (11 courses) than at the \bachelor level (5 courses). Only two universities (\UL and \UU) offer courses in this KA at both levels.

There are five courses that can be considered signature; three have the same name as the KA and are taught at \OU, \RU, and \UT; the \secure\ course at \UL which has a clear focus on system and software design; and the \SSSec\ course at \UL, which is an advanced version of the former. Out of the various courses, only two are offered as seminars: \SV\ (\UT) and \SSSec\ (\UL).


When analysing the most common keywords (\autoref{tab:top-words-2}, third column), we observe how --- after the words that identify the KA itself and the typical word `system' --- we find the generic term `analysis', the keyword `vulnerability' occurs, one of the key terms in the field. The following keywords based on frequency highlight basic facets of the KA: `secure', `programming', `software development', and `programming language'.


\subsection{SE Economics}
The KA \economics is not popular; only nine courses have a focus on this KA. These are taught across seven universities, although one (\UU) has a prevalence with three courses. Two courses are taught at the \bachelor level, the other seven are at the \master level. Furthermore, two of the \master courses are seminars, indicating that this is a topic that is still less mature when it comes to its teaching in higher education (although this is a well established topic in industry). Six of the nice courses are part of a SE-specific track.

\finding{\economics is the least popular KA in our analysis. This aligns with the early observations by Barry Boehm~\cite{boehm2000software} and the limited embedding in standard curricula; the 2014 ACM SE curriculum recommendations indicate a total of \textit{only} eight hours for this KA~\cite{ardis2015se}.}


Analysing the most frequent keywords in the course descriptions (\autoref{tab:top-words-2}, fourth column), we see `software' appearing as the top, unsurprisingly. The following keywords, with the exception of `development', are more interesting: they refer to `product', `management', `software product', `software ecosystem', and `business'. These have clear links to research domains that are highly linked to industrial needs, including software production~\cite{xu2007concepts}, software business~\cite{cusumano2008changing}, and software ecosystems~\cite{manikas2013software}.


\subsection{Verification}
The KA \verif, which we added to those from the SWEBOK because of its prominence in the SE landscape in the Netherlands, totals 28 courses taught by all the ten universities under analysis. The courses regarding verification are mostly taught in \master programmes (18 courses), with ten courses at the \bachelor level. It is worth noting that 8/10 universities include \verif courses at the \bachelor level, with the exceptions of \OU and \UvA. All ten universities offer at least one course at the \master level. Almost half of the courses (14) are part of a \master or a track that is SE specific. Three courses are given as seminars: two at \UT, one at \UL.

We do not provide an analysis of signature courses because we have defined \verif as a grouping for the courses that focus on the wide range of formal techniques that can be used to provide guarantees on the correctness of software systems. 




Concerning the most frequently occurring keywords (\autoref{tab:top-words-2}, last column), the most prominent ones are `program' and `system', which denote the element that is being verified. These are followed by the generic `technique', and the \SE specific `software' against which properties are verified. Other important keywords are `security' and `model checking', showing one of the key goals of verification and a fundamental technique for verifying software and systems.

\subsection{PL Design}
There are 24 courses in the KA \pldesign; half at the \bachelor level, half at the \master level. The \bachelor courses are offered by nine universities (except for \TUe), while the twelve \master-level courses are offered by only six universities, with \RU and \UU offering four and three courses each. This highlights the different emphasis put on this KA, especially at the \master level. Only nine courses are taught in SE-specific tracks or programs.


Just like for \verif, we do not provide signature courses. Nonetheless, we observe that six courses focus on compiler construction (\RU, \RUG, \TUD, \UU, \UvA, \VU), four courses address concepts of programming languages (\OU, \TUD, \UL, \UU), and two courses concern domain-specific languages (\UU, \TUe).


\begin{figure}[pos=h]
    \centering
    \includegraphics[width=0.9\columnwidth]{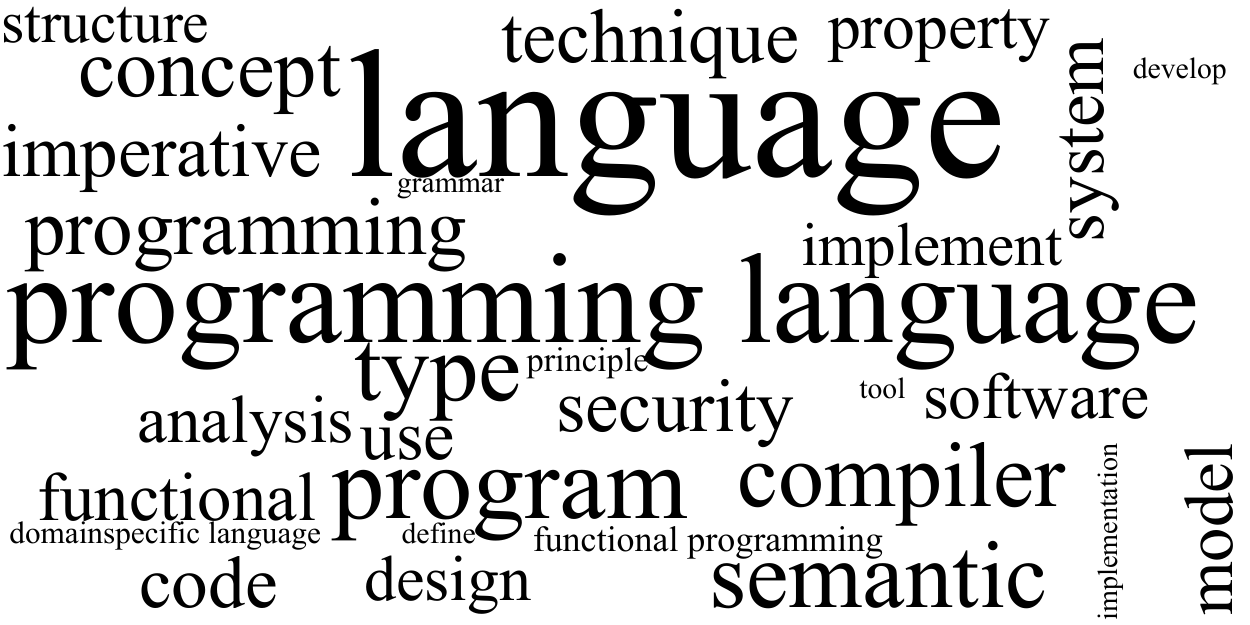} 
    \caption{Word cloud visualisation of the topics touched upon by the courses that relate to the KA \pldesign. }
    \label{fig:topics-pldesign}
\end{figure}

When analysing the most recurring keywords in \autoref{fig:topics-pldesign}, after the obvious `language', `programming language', and `program', we find `type', which refers to the study of type theory, a fundamental pillar in \pldesign~\cite{LangDesign2017}. This is followed by `semantic', as language semantics is a fundamental topic. The word cloud highlights then `compiler' (\pldesign informs compiler construction) and `concept' (but this is originating from the name of some courses).

\section{Correlations ($\mathbf{RQ_2}$)}
\label{sec:rq2}

\begin{figure}[pos=h]
    \centering
    \includegraphics[width=\linewidth]{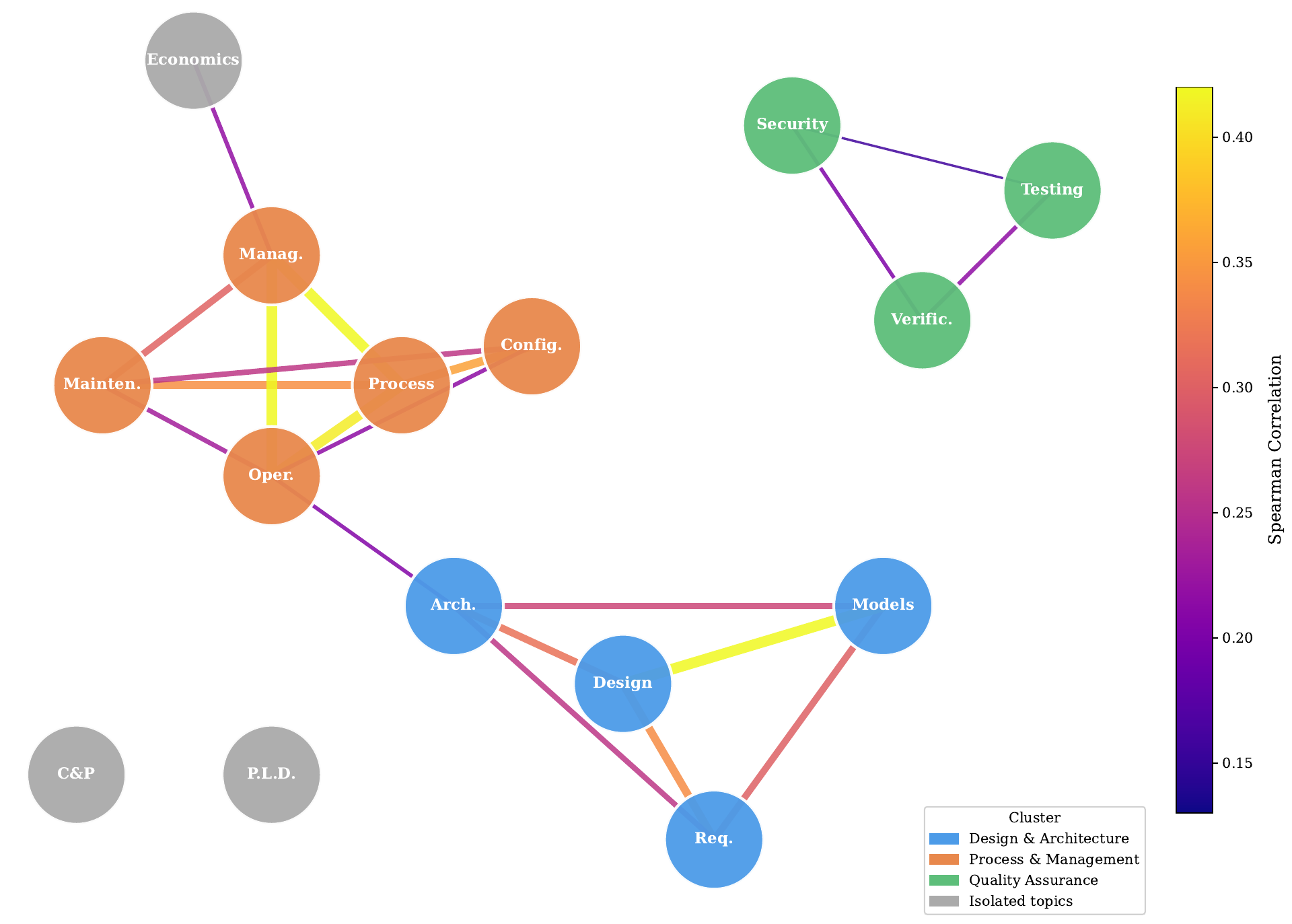}
    \caption{\revtwo{Correlations between the KAs in the analysed courses, and course clusters. Some KA names are shortened: Req: Requirements; Arch: Architecture; Constr: Construction \& Programming; Oper: SE Operations; Mainten: Maintenance; Config: Configuration management; Manag: SE management; Verific: Verification. P.L.D.: Programming language design, C\&P: Construction and Programming.}}
    \label{fig:course-network}
\end{figure}

After having analysed the details per KA, we now address \textit{RQ}$_2$ by studying the correlations between the KAs. We have determined these by calculating Spearman correlation (we could not assume the data are normally distributed) based on the KAs that were ticked for each course. The raw results are in the online appendix; an overview of the strongest positive correlations is provided in Fig.~\ref{fig:course-network}, with the line width and brightness denoting the strength of the correlation between two KAs. Below, we discuss the main clusters of courses based on the existence of stronger correlations between them (conventional interpretations for correlation~\cite{evans1996straightforward}: negligible when $\rho\in[0.00,0.20)$, weak when $\rho\in[0.20,0.40)$, moderate when $\rho\in[0.40,0.60)$, strong  when $\rho\in[0.60,0.80)$, and very strong when $\rho\in[0.80,1.00]$). \rev{Note that, in this context, even weak correlations are indicators of KAs that co-occur often: for a strong correlation to exist, most courses shall cover \textit{both} KA1 and KA2, and this is unlikely, especially in specialized courses that cover a single KA. All the correlation values are shown in the appendix at the end of this paper, in \autoref{tab:correlations}.}

\subsection{Requirements, Architecture, Design, SE Models}
\label{sec:cluster1}
The KA \req has weak correlations with the KAs \des (9/21 \req courses also cover \des, $\rho=0.34$), \modeling (9/21, $\rho=0.30$); and  \arch (7/21, $\rho=0.26$). This can be explained by the fact that architecture and design are generally seen as phases that follow RE in software development~\cite{nuseibeh2001weaving}. The co-occurrence of \modeling indicates that modelling is an important facet of RE. This is not only testified by the rich literature on requirements modelling, but also by the fact that professional training organisations such as the International Requirements Engineering Board (IREB) do offer certifications on the KA in their advanced level materials~\cite{van2016handbook}.

Within the seven signature courses, a different situation is portrayed: none of them teaches also \arch, one of them focuses on \des, and two of them cover \modeling. This shows that, while the \req KA is often co-taught with other topics in general courses, the more specific \req courses are less focused on the relationships with other areas of SE. 

In addition to the weak correlation with the above-mentioned KA \req, the KA \arch has a weak correlation with \des (9/22 courses, $\rho=0.31$) and with \modeling (9/22, $\rho=0.27$). Both terms appear in the second column of \autoref{tab:top-words-1}. This is not surprising, as both \des and \arch focus on design abstractions, with the main difference that \arch focuses on an additional level that goes beyond the organisation of code and, as explained by Rozanski and Woods~\cite{rozanski2012software}, that bridges the gap between the problem and solution spaces.

The KA \des has a moderate correlation with \modeling (13/24 courses, $\rho=0.42$), in addition to the above-mentioned weak correlation with \req (9/24, $\rho=0.34$) and \arch (9/24, $\rho=0.31$). The correlation with \modeling is unsurprising, as software design is often taught by means of the UML modelling language.

The KA \modeling moderate correlation with \des (13/28 courses, $\rho=0.43$) is witnessed by the existence of books that explain software design via modelling, such as that by Gomaa~\cite{gomaa2011software}. The weak correlations with \req (9/28, $\rho=0.30$) and with \arch (9/28, $\rho=0.27$) show the typical positioning of modelling as a communication tool that informs the activities before the construction of software systems, in addition to the model-driven SE paradigm~\cite{brambilla2017model}.

\finding{\req, \arch, \des and \modeling are a tightly coupled set of KAs that represent the process for detailing software systems: stakeholders' requirements $\Rightarrow$  system-level architecture $\Rightarrow$ detailed design decisions, often using models to represent those aspects.}

\subsection{Testing, Security, Verification}
\label{sec:cluster2}
The cluster of the three KAs refers to well-known knowledge in computing education~\cite{garousi2020software}: in order to make a system more secure, verification and testing are key activities, with the former concerned with ensuring security in provable manners, and the latter focused on providing probabilistic evidence when full coverage via formal models is not feasible. Despite the three KA of this cluster are linked, we observe that their correlation is overall \textit{more loose than the other clusters} \rev{(hence, we do not identify a finding)}, displaying that, albeit commonalities shared by the three KAs, the topics are often addressed in silos. 

The \test KA (19 courses) has weak correlations with \verif (7 shared courses, $\rho=0.21$) and a negligible-to-weak one with \security (4 shared courses, $\rho=0.15$). Furthermore, \security (16 courses) has a weak correlation with \verif (28 courses, 6 shared with \security, $\rho=0.21$).  

\subsection{SE Operations, SE Management, SE Process, Maintenance, Configuration Management}
\label{sec:cluster3}
The final KA-based cluster of courses concerns different facets of the DevOps paradigm. 
This cluster results to be only marginally linked to the other clusters, with only few weak correlations between \semgmt and \economics ($\rho=0.21$) and \ops and \arch ($\rho=0.20$). 

Inside the cluster, we observe that the topics are the most highly cohesive across all clusters, with moderate correlations being between \seproc and \ops ($\rho=0.41$), \seproc and \semgmt ($\rho=0.42$), and \semgmt and \ops ($\rho=0.42$). 

Overall, by studying the correlations within this cluster, we can conclude that the topics related to DevOps are seldom taught singularly across all considered universities. We conjecture that this could be either due to (i) the impossibility of treating the the theoretical concepts of the KA alone without considering the other KAs, (ii) a general lack of depth with which the topics are treated, leading to the need of considering multiple KA to fill in the content of courses, or (iii) the dissection of a higher concept considered in the course (e.g., DevOps) into finer-grained KA by the SWEBOK. 

\finding{Because of their interconnectedness, DevOps topics are rarely taught singularly (e.g., via signature courses), but rather covered in courses that touch upon multiple related KAs: \ops, \semgmt, \seproc, \maint, \conf.}

\subsection{Lone Riders: Construction \& Programming, PL Design, and SE economics}
Three KAs appear to be almost completely isolated from the others, namely \economics, \pldesign, and \cnp, although potentially due to different reasons.

In the case of \cnp, the topic treated results to be of a peculiar nature, leading to it need to be considered independently to other KAs. More specifically, a key kind of course in \cnp is the introductory programming course of many SE curricula, which lays the basis on which students can then follow up upon by considering more specialised topics. Therefore, as introductory course, more advanced SE topics are very seldom considered in depth during courses covering this core KA.

Regarding \pldesign, instead, its detachment from other KAs might more be due to the theoretical and formal nature of the topic considered, leading it to be only marginally considered in isolation during SE-centric educational curricula.

Finally, the \economics course is the only that presents a weak correlation to other KA, namely \semgmt ($\rho=0.21$). As for the other KA considered above in this section, we conjecture that the peculiar nature of the topic, which focuses more on economic reasoning rather than technical SE aspects, leads the topic to be often treated in specialised courses.

\section{University Educational Foci ($\mathbf{RQ_3}$)}\label{sec:rq3}
\begin{figure}[pos=h]
    \centering
    \includegraphics[width=\linewidth]{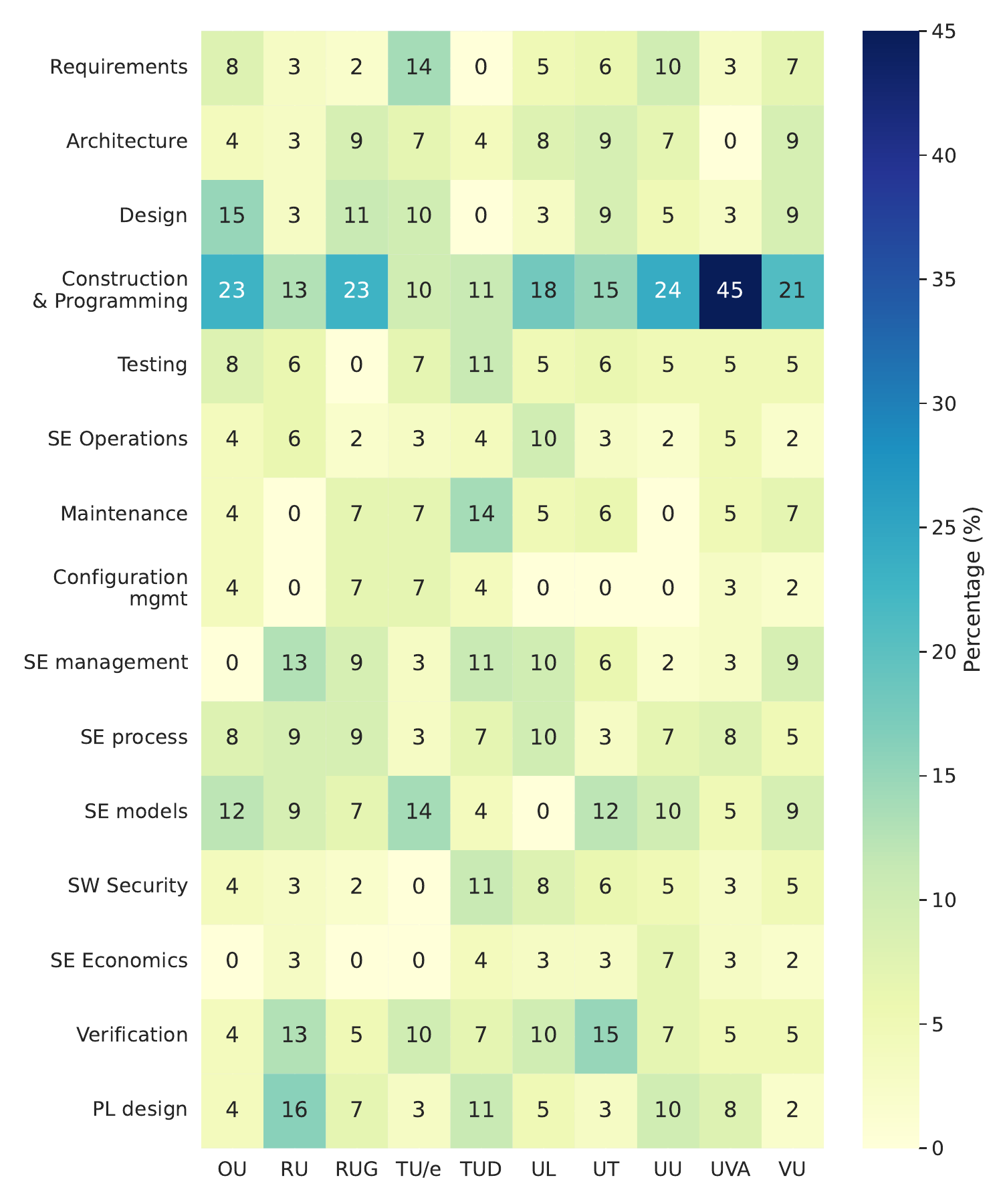}
    \caption{\rev{Heatmap across KAs and universities: percentage of courses covering a specific KA within a given university.}}
    \label{fig:foci}
\end{figure}

In this section, we address $RQ_3$ by reporting the results related to the educational focus provided at the ten Dutch universities considered.
The distribution of the overall SE education across KAs at the different universities is depicted in Figure~\ref{fig:foci}.
The figure shows that, while several commonalities can be identified across the different universities, some specificities exist too. 

In terms of commonalities, most universities result to provide a high number of courses providing a \cnp component, with the \UvA providing the highest recurrence of such topic across its SE educational offering. Rather than a specific pedagogical focus on such KA, we deem this high recurrence due to the course \emph{Introduction to programming}\textbf{} provided at the \UvA, which is duplicated across multiple curricula at that university.
Overall, the high frequency of the \cnp KA across universities can could be explained by the foundational nature of such KA, especially prevalent at the \bachelor level in our dataset.

Other commonalities emerge for less recurrent KAs, such as \arch, \verif, and \economics. The specialised and advanced nature of such KAs results in their overall low recurrence across all universities, with the exception of \UT, that seems to provide a higher rate of specialisation courses on verification topics.

Some KA result instead to be treated with different intensity at the universities considered. For example, the presence of the \security KA, while never constituting a considerable portion of SE curricula, varies from a limited fraction of courses (\RU) to a notable portion of the educational offerings (\TUD). Similarly, a notable portion of courses at \RU focus on the \pldesign KA, while the topic results to be less recurrent at other institutions (e.g., \UT and \VU).

Some KA result to be absent at certain institution. Rather than the complete lack of such KAs, we associate such finding to a limited coverage of related topics at the considered institutions, that led to their low representation in the course syllabi, and hence were not included \textit{via} our data collection process (see Section~\ref{sec:research_process}). 

Overall, we note that all educational contexts provide an extensive and heterogenous range of educational offers, with no institution delivering a narrow and specialised focus on a specific SE topic. The vast range of topics covered in the educational curricula of the considered universities, jointly with the absence of specific pedagogical foci, might be a pointer to the multifaceted skills required to train modern software engineers, which are necessary regardless of specific SE specialisations provided at the universities (e.g., the \textit{Software Engineering and Green IT} \master track at the \VU).

\revtwo{As further note, there are significant global trends that may reshape the SE educational landscape. The popularization of large language models (LLMs) alone might have a noticeable ripple effect on the educational offerings of universities, from the introduction of new standalone, specialized Bachelor programs in fields such as Generative AI, to the restructuring and redesign of existing SE courses and university curricula. We acknowledge that the educational snapshot of the current state of SE education in the Netherlands might soon be exposed to a some degree of restructuring. Additionally, while our data focuses on computer science programs because SE courses are predominantly embedded there in the Netherlands, it would be interesting to explore degrees beyond ``classical'' computer science, such as in AI or data science, and compare the results.}

\finding{While minor differences exist for specific KAs (e.g., \security and \pldesign), the analyzed universities provide a high degree of commonality in terms of the covered KAs, indicating that the studied KAs are necessary traits of modern software engineers. We expect more noticeable differences in terms of the specialization focus, e.g., sustainability or artificial intelligence. These aspects are, however, not covered by the SWEBOK, the theoretical lens of our analysis. \revtwo{The rapid adoption of LLMs might soon have a noticeable ripple effect on SE education, leading to the restructuring and redesign of existing courses and university curricula.}}

\section{Related Work}
\label{sec:rel_work}
Lee et al.~\cite{lee2011change} showed that Taiwan faces significant challenges in SE education, with issues in both the quantity and quality of prepared graduates. 
Despite a surplus of IT-related students, many lack core competencies. This is attributed to an overemphasis on introductory courses and neglect of advanced topics like software quality, modelling, and domain knowledge. 
The  Consortium of Taiwan has addressed these gaps by introducing a module-oriented curriculum inspired by SWEBOK, leading to increased course offerings and enrolment. 
However, persistent dysfunctions, such as avoidance of process and insufficient attention to practical domain applications, still hinder the alignment of education with industry needs.
Our research extends these efforts by analysing SE education in the Netherlands, focusing on the breadth and co-occurrence of KAs across graduate and undergraduate courses. 
Unlike previous studies, we analyse interdependencies among KAs and assess variations in educational focus across universities. 
This approach offers a comparative perspective to identify unique strengths and gaps in preparing graduates for SE challenges. Despite the lack of ambition to country-wide curriculum revision, we offer comparable insights gained from comparing existing SE courses in the Netherlands, and aligning them with SWEBOK chapters.

In their systematic review, Tenhunen et al.~\cite{ tenhunen2023systematic} analysed 127 papers related to SE capstone courses (2007--2022). 
They found that capstone courses in computer science and SE programs aim to prepare students for professional life by combining technical skills with essential soft skills such as teamwork, communication, and project management. 
They identified common features of SE capstone courses, such as semester-long duration, team sizes of 4--5 students, and the use of real-world projects, often with external clients. 
While these courses effectively cover software development life-cycle stages, gaps remain in areas like software maintenance and the integration of continuous peer assessment. 
While this systematic review provides a comprehensive understanding of current SE capstone courses, our research sets itself apart by focusing specifically on the educational landscape within the Netherlands.

Huang et al.~\cite{huang2021research} presented a tertiary study that synthesised and classified 26 SLRs published between 2004 and 2019. 
They noted that while several individual SLRs have focused on specific aspects of SE education, none have provided a broad, high-level overview of the entire SE education research landscape. 
By offering insights into the frequently applied methods and tools, as well as the recurring themes and issues in SE education, they provided a broader and more holistic view of the field. 
Additionally, they identified key challenges in SE education that have emerged from these studies which can help to inform future research and development in SE pedagogy.

Malik and Zafar~\cite{malik2012systematic} conducted a mapping study focusing on improving SE education practices. 
This study provided an overview of the literature addressing SE education advancements, identifying and categorising 70 primary studies published according to a well-defined SE educational framework.
The authors highlighted that inadequate curricula for SE remain a critical risk, emphasising the need for curricula that align with industry demands and equip students with the necessary skills, knowledge, and expertise.
The study mapped primary studies to reference curricula such as SWEBOK, identifying gaps in these frameworks that could benefit from revision to better reflect diverse aspects of SE. 
Additionally, they analysed publication trends, noting that conferences like CSEE\&T and regions like the USA and China lead in SE educational research.
The authors proposed that their findings can inform curriculum refinement, align academic practices with industry needs for practitioners, and inspire new research directions for researchers.

Qadir and Usman~\cite{qadir2011software} presented a systematic mapping study to synthesise and aggregate literature on SE curriculum design, revision, and assessment, providing a comprehensive overview of efforts reported in conferences like CSEE\&T, FIE, and EDUCON.
Their study highlighted the historical evolution of SE as a discipline, emphasising the transition from its early conception as a subfield of Computer Science to a mature academic field with stand alone programs at \bachelor and \master levels. 
The authors noted that contributions related to course-level improvements far outnumber those focused on programme-level enhancements, signalling a need for greater emphasis on holistic curriculum development.
They underlined key frameworks such as SWEBOK, SE 2004, and GSwE2009, which have guided SE curriculum advancements, and also identified a growing need to align curricula with industry trends and demands.
Our research builds on this work by exploring the state of SE curricula within the Netherlands, focusing on the co-occurrence and integration of knowledge areas across various academic institutions.
Unlike this mapping study, which primarily categorises contributions by quantity and type, our approach evaluates the interdependencies of curriculum components and their practical alignment with industry requirements, offering actionable insights into the broader landscape of SE education.

Liargkovas et al.~\cite{liargkovas2021software} investigated the gap between software practitioners' education, as outlined in the IEEE/ACM SE2014 SEEK, and industrial needs, using Wikipedia articles cited in Stack Overflow posts as a proxy for developers’ real-world informational needs. 
Their study revealed that while SEEK adequately covers foundational topics like computer science fundamentals, software design, and mathematical concepts, it lacks sufficient emphasis on practical areas such as the World Wide Web, testing, security, and soft skills. 
The authors identified significant misalignments between the SEEK curriculum and industry requirements, suggesting that future curriculum models should integrate industry-driven topics (e.g., software testing, design, and security) and soft skills (e.g., teamwork, collaboration, and project management) to better prepare students for professional challenges. 
Our research complements these findings by focusing on curriculum design within the Netherlands, evaluating the integration and co-occurrence of knowledge areas across universities to identify unique gaps and opportunities for alignment with industrial demands.

\rev{Some studies analyzed the academia-industry gap through surveys with SE graduates. Lethbridge~\cite{lethbridge2002knowledge}'s seminal work was the first one to explore the gap in terms of which topics were considered important, were learned in education, and/or learned on the job. Lethbridge's instrument was later adapted by Kitchenham \textit{et al.}~\cite{kitchenham2005investigation}, who conducted a similar survey with thirty graduates of four universities in the United Kingdom. Our study focuses on analyzing the course descriptions, rather than surveying graduates, although we reckon that triangulating with graduates' perception would be interesting.}

A systematic mapping study on Software Engineering for Sustainability (SE4S) provided an updated and in-depth overview of the field, investigating recent contributions, knowledge areas, and research facets~\cite{penzenstadler2014systematic}. 
They offer a research map that identified trends, tools, methods, and frameworks.
The findings revealed that while SE4S has grown significantly, its impact remains concentrated in a few knowledge areas, highlighting gaps that require further exploration.
The authors emphasised the need for a roadmap to guide future SE4S research and development. 
This study aligns with broader efforts to adapt SE practices to address sustainability challenges, offering insights for aligning curricula and research agendas with emerging societal needs.

In contrast to prior research, including studies focused more narrowly on evidence-based SE practices~\cite{kitchenham2007large, 10.1145/3477535}, and those offering a broader, more holistic view of SE education~\cite{huang2021research, malik2012systematic}, our study serves as a reference point for academics and practitioners looking to understand SE education in the Netherlands. 
By examining the co-occurrence and integration of KAs across undergraduate and graduate programs, we provide a nuanced perspective on how Dutch institutions align their curricula with industry requirements and emerging trends in SE. 
This research not only highlights gaps and opportunities within the Netherlands’ educational landscape but also offers actionable insights into curriculum design and interdependencies among KAs, bridging the divide between academic preparation and professional practice.

\section{Discussion}\label{sec:discussion}
We analyse the findings of our study in relation to the research questions, highlighting key insights and their significance.
We begin by addressing the research questions and interpreting the results in the context of our study.
Next, we discuss potential threats to validity and limitations, acknowledging factors that may have influenced our outcomes. 
Finally, we explore the broader implications of our findings for SE education, providing insights that may inform future research and practice.

\subsection{Answers to the research questions}

$RQ_1$: \textit{What are the topics studied in SE Courses?}\\
We found that the various KAs are generally covered to a reasonably comparable extent (see \autoref{fig:ka_occurrence}). While the KA \cnp is the most frequent (74 courses, while the second most common, \modeling, has 28 courses), this can easily be explained by the fact that programming is a basic skill for both computer science and \SE curricula, also demonstrated by the fact that 62/74 courses in the \cnp KA are at the \bachelor level. Some differences can be seen when comparing KAs across \bachelor and \master: for example, \maint is mostly covered by \master courses (18/19), while other KAs are perfectly balanced, like \modeling (14 per level), \des (12 per level), and \pldesign (12 per level). The presence of a high number of courses covering the KA \verif is a remarkable finding, which is in line with the research tradition in the Netherlands. A couple KAs have only nine courses (less than one per university); these are \conf and \economics. The word clouds included in \autoref{sec:categories} provide a more in-depth analysis for the reader interested in learning the most frequent topics mentioned in the course descriptions.

\vspace{0.2cm}
\noindent $RQ_2$: \textit{What are the (co-)occurrences among studied topics and the educational context?}\\
We conducted a correlation analysis of the KAs covered by the courses (see \autoref{fig:course-network}). The results highlight three clusters of KAs that are tightly coupled (i.e., courses often cover more than one of such KAs), explained in \autoref{sec:cluster1}--\autoref{sec:cluster3}: (i) \SE activities that pertain to the requirements analysis and design (\req, \arch, \des, \modeling); (ii) techniques and tools for quality assurance, including security (\test, \security, \verif); and (iii) process-oriented and DevOps-related topics (\ops, \semgmt, \seproc, \maint, \conf). The analysis also revealed some isolated KAs that are generally taught on their own: these are \cnp, \pldesign, and \economics. The former two can be explained by the fact these topics are part of general computer science programs; for \economics, we speculate this is occurs due to the existence of very specialised courses that focus on the business and managerial aspects of \SE.

\vspace{0.2cm}
\noindent $RQ_3$: \textit{Do universities differ in terms of education foci?}
Our analysis of the number of courses covering each KA at the various institutions (see \autoref{fig:foci}) does not reveal significant discrepancies: Dutch universities seem to cover all the analysed KAs via their courses. The most remarkable outlier is the high number of  courses on \cnp at \UvA, but this is explained by the existence of numerous variants of  programming courses that are offered to various programs across the university. Smaller deviations exist, but they simply reflect specificities of some universities, often originating from their research foci.

\subsection{Threats to Validity and Limitations}
In this section, we discuss the most relevant threats to validity that might have influenced our research, by considering common pitfalls and suggestions to consider threats in empirical studies~\cite{lago2024threats}. Additionally, we document the limitations our research results entail. 

\subsubsection{Internal Validity}
A potential internal threat to our results is constituted by the labelling strategy utilised to code the courses according to the predefined KA. To mitigate this threat, we relied on a crowdsourcing strategy by involving educators of all universities to tag their own courses (see Phase 4 of our research method in Section~\ref{sec:research_process}). The final tags resulting from the homogenisation strategy (Phase 5) were then revised again by the original annotators, in order to further mitigate internal threats arising from the homogenisation process. \rev{The tagging reached \textit{substantial} agreement $\kappa=0.718$, although we acknowledge that a higher degree of agreement would be desirable.}

In order to mitigate potential threats related to the data analysis, only simple summary statistics are utilised to present the results, while reporting as much as possible the raw data on which our results are based (e.g., the number of courses considering a specific KA).

\subsubsection{Construct Validity}
\revtwo{An inherent threat to the validity of our results lies in the predefined Knowledge Areas (KAs) used for data collection. Specifically, the KAs outlined in the SWEBOK guide might not fully capture the nuanced topics of the analyzed software engineering university courses. Relying on an \textit{a priori} defined catalog of KAs was deemed however a necessary tradeoff between construct validity and internal validity. While such research design choice potentially sacrificed some expressiveness and simplified the studied construct, it helped mitigate subjective bias during the data analysis process. As overarching mitigation strategy of this threat, we designed a tailored research process to refine and adapt the SWEBOK KAs specifically for our study's context (see Phase 3 of the research method in Section~\ref{sec:research_process}).}

\revtwo{Regarding a further simplification of the observed construct that might have influenced our results, in our study we do not go into the details on study program structures, e.g., by considering mandatory versus optional courses, study credits associated to courses, prerequisite course ordering, nature of the thesis project, assigned grades, and student satisfaction questionnaires. While considering these characteristics would have yielded to a more nuanced understanding of the educational landscape, doing so for all 207 courses would have been prohibitively challenging, and opening to threats such as data deluge and analysis paralysis. The design choice of not considering such information was hence a necessary tradeoff between construct validity and both internal validity and study feasibility. This limitation can be addressed in future work by designing a study specifically tailored to focus on these more granular facets.}

Finally, another potential threat to construct validity regards the original source of data of our study, namely the syllabi of the courses considered. In fact, the content of the silabi, which are usually drafted before the start of courses, might deviate from the actual content of the live lecture. Using the syllabi as original data source entailed a trade-off between construct and external validity, which allows us to consider a rather high number of courses, namely 207, that could not be considered if another type or research method, e.g., interviews with educators, would be adopted. To mitigate this threat, researchers label exclusively syllabi of their own university, resorting to contacting the educator in charge of the course whenever the information on the syllabi is deemed insufficient.

\subsubsection{External Validity}
The results presented in this study are by definition bound to the specific geo-educational context considered, namely SE higher education in the Netherlands. While the results and conclusions might be applicable to different extents to other countries, we cannot not claim the generalisability of the results. While the implications presented in Section~\ref{sec:implications} might be corroborated by considering similar pedagogical contexts to the one chosen for this study, further research is needed to prove such assertion.

At the intersection of external and construct validity, we warn interested readers and researchers that the research process, keywords, and resulting KA might be applicable and replicable only in similar educational and societal contexts, namely SE university courses with frontal lectures in Western countries. Further research should be conducted to study if, and in affirmative case to what extent, the research process and intermediate research artefacts can be transposed to different pedagogical contexts and paradigms such as peer learning. 

\subsubsection{Limitations}
As limitations to our work, due to the volume of courses considered, we are not able to study in depth many nuanced facets of the courses considered, such as topics covered within each KA, student grades, and number of educators involved in each course among others. Therefore, the results and conclusions presented in this study need to be interpreted exclusively as a higher level overview of the educational offerings of SE universities in the Netherlands. Numerous related concepts, such as student perceptions, impediments encountered by educators, \rev{graduate employability, industry expectations, and} effectiveness of different pedagogical components, and variation of the educational landscape in time, remain uncovered with this research and are left up for future work.

\subsection{Implications for SE education}
\label{sec:implications}
The findings suggest actionable insights that can be used to  enhance and diversify \SE curricula. 

\smallskip

\noindent\textit{Integrative courses on software design activities.} The pairwise correlations between the KAs \req, \des, \arch, and \modeling indicate that there may be an opportunity to foster a more integrated teaching approach across these KAs. 
Educators could consider designing broader courses that allow students to engage with these KAs holistically, with an emphasis on how they inform each other throughout the software development lifecycle. This could be done via the creation of modules that consist of sub-modules, each focusing on one specific KA, with an integrative project where the students would learn about the complex dependencies across the KAs via exercising their learned knowledge (and techniques) on a specific case.

\smallskip

\noindent\textit{Software security via testing and verification.}
Similarly, the correlations among \test, \security, and \verif provide another opportunity for educators to create integrated learning experiences. 
Given the increasing importance of security and reliability in software systems, educators could develop projects or case studies that require students to apply verification methods and testing techniques to ensure security in software systems. 
This could be particularly effective in advanced \master courses, where students can explore the theoretical underpinnings of these KAs and their application in practice. 
Additionally, offering seminars or project-based courses focused on security and verification could help students gain hands-on experience with state-of-the-art tools and methodologies.

\smallskip

\noindent\textit{Bridging the \economics island.}
The KA \economics is not only under-represented in the analysed SE curricula, but it is also isolated from the rest. 
As this area is highly relevant to the industry, and for students to comprehend the business aspects of software, educators should consider expanding the scope of \SE programs by integrating courses that address the economic aspects of software development, including business models, software ecosystems, and sustainability. 
Our analysis reveals that these topics, when taught, are more often embedded at the \master level; educators could consider introducing them earlier in the curriculum to better prepare students for the job market.
Interdisciplinary courses combining SE with business, economics, and entrepreneurship could be particularly valuable in this regard. 



\section{Summary and Future Work}\label{sec:summary}
This paper provided a systematic analysis of the \SE higher education landscape in the Netherlands, through an analysis of 207 courses offered by ten universities. We have identified that ($RQ_1$) the KA \cnp is unavoidably the most frequent, followed by a quite balanced distribution across the other KAs, with the exception of \conf and \economics. Our analysis of co-occurrences ($RQ_2$) revealed clusters of KAs that are often co-taught, regarding analysis and design, quality assurance, and SE process and DevOps. Finally, we could not find significant differences in the distribution of the KA coverage across universities ($RQ_3$). 

Our findings point to a need for further exploration into the teaching of \economics, a less represented KA that appears to be highly relevant to industry but still under-emphasised in academic settings. 
Researchers could investigate how to better integrate economic principles such as software production models, business ecosystems, and sustainability into SE curricula, especially at the \bachelor level. 
This could involve developing interdisciplinary courses or modules that combine SE with business and entrepreneurship.

The findings also suggest several important areas for future research, particularly in the context of the correlations between different KAs and their integration into SE education. 
One significant observation is the strong correlation between the KAs of \req, \des, \arch, and \modeling. 
This highlights the necessity for further research on how these KAs interact in both academic curricula and industry practices, with a focus on understanding the dependencies and synergies between them. 
Researchers should explore how these correlations can be leveraged to create more integrated learning modules or courses that bridge gaps between foundational concepts like \req and advanced topics like \arch and \des. 
The strong ties between \req and \des, as well as the role of \modeling as communication tools in the development lifecycle, suggest that more research is needed to understand how these KAs can be effectively taught together, potentially through model-driven SE paradigms.

Moreover, the cluster involving \test, \security, and \verif offers an opportunity to explore the interconnected nature of these KAs in the context of ensuring software quality and security. 
Researchers could investigate how the principles of verification and testing can be aligned with security practices in educational settings, particularly through cross-KA projects that allow students to apply formal verification methods, testing strategies, and security principles simultaneously. 
This would not only enrich students' practical skills but also enhance their understanding of the complexities involved in developing secure and reliable software systems. 
Research on the alignment between \verif and \pldesign is also promising, as both fields share formal mathematical foundations and could benefit from joint curricula focusing on formal methods, type systems, and language semantics.

\rev{A useful direction would be extending the study by considering the ordering of the courses and possible pre-requisites, in an attempt to study the progression across KAs within specific study paths, besides the correlations. Furthermore, there are important trends that are affecting computing education: while the SWEBOK is a robust resource for the mapping, it is rather conservative and does not cover emerging areas like Generative AI~\cite{daun2023chatgpt}, Quantum computing~\cite{murillo2025quantum}, sustainability~\cite{moreira2025roadmap}, etc. Future studies could take these topics into account as well.}

\revtwo{As future work, we plan to replicate our nationwide study within the Italian software engineering higher education system, where one of the authors is currently based. Conducting a study of this scale demands several prerequisites, including established institutional partnerships, deep familiarity with the local educational system, and the cooperation of at least one educator from each participating university. While meeting all these criteria simultaneously is challenging, we highly encourage researchers to repeat our study in their national context to uncover regional similarities and differences. Furthermore, we invite international comparative studies, though scholars have eloquently noted the substantial challenges such cross-border efforts~entail~\cite{teichler2014opportunities}.}

\section*{Acknowledgments}
We dedicate this work in loving memory of Bastiaan Heeren, who passed away while this work was being completed. His passion and dedication to software engineering education will not be forgotten, and we are proud and honored to have had the opportunity to collaborate with him. We would also like to thank all those who contributed to and helped with the collection and verification of the data, especially Marcello Bonsangue (University of Leiden), Andrea Capiluppi (University of Groningen), Patricia Lago (Vrije Universiteit Amsterdam), Ana Oprescu (University of Amsterdam), Annibale Panichella (Delft University of Technology).





\bibliographystyle{cas-model2-names}
\bibliography{bibliography.bib}

\onecolumn

\clearpage 

\appendix 
\section{Correlations between the KAs}

\begin{table}[pos=h]
\centering
\caption{Correlations between the KAs (via Spearman). The color coding is based on the standard interpretation by Evans~\cite{evans1996straightforward}: very strong ($|r_s| \ge 0.8$): \colorbox{vstrong_pos}{\textcolor{white}{Dark Green}}; strong ($0.6 \le |r_s| < 0.8$): \colorbox{strong_pos}{Medium Green}; moderate ($0.4 \le |r_s| < 0.6$): \colorbox{moderate_pos}{Light Green}; weak ($0.2 \le |r_s| < 0.4$): \colorbox{weak_pos}{Pale Green}; negligible ($|r_s| < 0.2$): \colorbox{negligible}{Pale Yellow}. For negatives, similar shading but in tones of red.}
\label{tab:correlations}
\begin{adjustbox}{width=\textwidth}
\small
\begin{tabular}{l|ccccccccccccccc}
\toprule
 & \rotatebox{90}{\req} & \rotatebox{90}{\arch} & \rotatebox{90}{\des} & \rotatebox{90}{\cnp} & \rotatebox{90}{\test} & \rotatebox{90}{\ops} & \rotatebox{90}{\maint} & \rotatebox{90}{\conf} & \rotatebox{90}{\semgmt} & \rotatebox{90}{\seproc} & \rotatebox{90}{\modeling} & \rotatebox{90}{\security} & \rotatebox{90}{\economics} & \rotatebox{90}{\verif} & \rotatebox{90}{\pldesign} \\
\midrule
\req & \cellcolor{vstrong_pos}\color{white}1.00 & \cellcolor{weak_pos}0.26 & \cellcolor{weak_pos}0.34 & \cellcolor{negligible}-0.11 & \cellcolor{negligible}0.01 & \cellcolor{negligible}0.03 & \cellcolor{negligible}-0.05 & \cellcolor{negligible}-0.07 & \cellcolor{negligible}0.09 & \cellcolor{negligible}0.13 & \cellcolor{weak_pos}0.30 & \cellcolor{negligible}-0.09 & \cellcolor{negligible}-0.07 & \cellcolor{negligible}-0.08 & \cellcolor{negligible}-0.12 \\ \hline
\arch & \cellcolor{weak_pos}0.26 & \cellcolor{vstrong_pos}\color{white}1.00 & \cellcolor{weak_pos}0.32 & \cellcolor{negligible}-0.19 & \cellcolor{negligible}-0.11 & \cellcolor{weak_pos}0.21 & \cellcolor{negligible}0.11 & \cellcolor{negligible}0.08 & \cellcolor{negligible}0.12 & \cellcolor{negligible}-0.08 & \cellcolor{weak_pos}0.28 & \cellcolor{negligible}-0.10 & \cellcolor{negligible}0.08 & \cellcolor{negligible}-0.14 & \cellcolor{negligible}-0.12 \\ \hline
\des & \cellcolor{weak_pos}0.34 & \cellcolor{weak_pos}0.32 & \cellcolor{vstrong_pos}\color{white}1.00 & \cellcolor{negligible}0.01 & \cellcolor{negligible}-0.01 & \cellcolor{negligible}0.07 & \cellcolor{negligible}0.04 & \cellcolor{negligible}0.14 & \cellcolor{negligible}0.06 & \cellcolor{negligible}0.14 & \cellcolor{moderate_pos}0.43 & \cellcolor{negligible}-0.10 & \cellcolor{negligible}-0.08 & \cellcolor{negligible}-0.14 & \cellcolor{negligible}-0.13 \\ \hline
\cnp & \cellcolor{negligible}-0.11 & \cellcolor{negligible}-0.19 & \cellcolor{negligible}0.01 & \cellcolor{vstrong_pos}\color{white}1.00 & \cellcolor{negligible}-0.14 & \cellcolor{negligible}-0.17 & \cellcolor{weak_neg}-0.20 & \cellcolor{negligible}-0.01 & \cellcolor{weak_neg}-0.24 & \cellcolor{negligible}-0.13 & \cellcolor{negligible}-0.03 & \cellcolor{negligible}-0.03 & \cellcolor{negligible}-0.11 & \cellcolor{weak_neg}-0.21 & \cellcolor{negligible}-0.08 \\ \hline
\test & \cellcolor{negligible}0.01 & \cellcolor{negligible}-0.11 & \cellcolor{negligible}-0.01 & \cellcolor{negligible}-0.14 & \cellcolor{vstrong_pos}\color{white}1.00 & \cellcolor{negligible}0.04 & \cellcolor{negligible}-0.04 & \cellcolor{negligible}0.01 & \cellcolor{negligible}-0.12 & \cellcolor{negligible}0.04 & \cellcolor{negligible}0.17 & \cellcolor{negligible}0.16 & \cellcolor{negligible}-0.07 & \cellcolor{weak_pos}0.22 & \cellcolor{negligible}-0.12 \\ \hline
\ops & \cellcolor{negligible}0.03 & \cellcolor{weak_pos}0.21 & \cellcolor{negligible}0.07 & \cellcolor{negligible}-0.17 & \cellcolor{negligible}0.04 & \cellcolor{vstrong_pos}\color{white}1.00 & \cellcolor{weak_pos}0.23 & \cellcolor{weak_pos}0.21 & \cellcolor{moderate_pos}0.42 & \cellcolor{moderate_pos}0.41 & \cellcolor{negligible}0.05 & \cellcolor{negligible}-0.08 & \cellcolor{negligible}0.03 & \cellcolor{negligible}-0.11 & \cellcolor{negligible}-0.10 \\ \hline
\maint & \cellcolor{negligible}-0.05 & \cellcolor{negligible}0.11 & \cellcolor{negligible}0.04 & \cellcolor{weak_neg}-0.20 & \cellcolor{negligible}-0.04 & \cellcolor{weak_pos}0.23 & \cellcolor{vstrong_pos}\color{white}1.00 & \cellcolor{weak_pos}0.26 & \cellcolor{weak_pos}0.30 & \cellcolor{weak_pos}0.34 & \cellcolor{negligible}-0.03 & \cellcolor{negligible}-0.03 & \cellcolor{negligible}0.10 & \cellcolor{negligible}-0.13 & \cellcolor{negligible}-0.12 \\ \hline
\conf & \cellcolor{negligible}-0.07 & \cellcolor{negligible}0.08 & \cellcolor{negligible}0.14 & \cellcolor{negligible}-0.01 & \cellcolor{negligible}0.01 & \cellcolor{weak_pos}0.21 & \cellcolor{weak_pos}0.26 & \cellcolor{vstrong_pos}\color{white}1.00 & \cellcolor{negligible}0.07 & \cellcolor{weak_pos}0.36 & \cellcolor{negligible}-0.02 & \cellcolor{negligible}-0.06 & \cellcolor{negligible}-0.05 & \cellcolor{negligible}-0.08 & \cellcolor{negligible}-0.08 \\ \hline
\semgmt & \cellcolor{negligible}0.09 & \cellcolor{negligible}0.12 & \cellcolor{negligible}0.06 & \cellcolor{weak_neg}-0.24 & \cellcolor{negligible}-0.12 & \cellcolor{moderate_pos}0.42 & \cellcolor{weak_pos}0.30 & \cellcolor{negligible}0.07 & \cellcolor{vstrong_pos}\color{white}1.00 & \cellcolor{moderate_pos}0.42 & \cellcolor{negligible}-0.05 & \cellcolor{negligible}-0.10 & \cellcolor{weak_pos}0.22 & \cellcolor{negligible}-0.14 & \cellcolor{negligible}-0.13 \\ \hline
\seproc & \cellcolor{negligible}0.13 & \cellcolor{negligible}-0.08 & \cellcolor{negligible}0.14 & \cellcolor{negligible}-0.13 & \cellcolor{negligible}0.04 & \cellcolor{moderate_pos}0.41 & \cellcolor{weak_pos}0.34 & \cellcolor{weak_pos}0.36 & \cellcolor{moderate_pos}0.42 & \cellcolor{vstrong_pos}\color{white}1.00 & \cellcolor{negligible}0.07 & \cellcolor{negligible}-0.05 & \cellcolor{negligible}0.14 & \cellcolor{negligible}-0.10 & \cellcolor{negligible}-0.13 \\ \hline
\modeling & \cellcolor{weak_pos}0.30 & \cellcolor{weak_pos}0.28 & \cellcolor{moderate_pos}0.43 & \cellcolor{negligible}-0.03 & \cellcolor{negligible}0.17 & \cellcolor{negligible}0.05 & \cellcolor{negligible}-0.03 & \cellcolor{negligible}-0.02 & \cellcolor{negligible}-0.05 & \cellcolor{negligible}0.07 & \cellcolor{vstrong_pos}\color{white}1.00 & \cellcolor{negligible}-0.06 & \cellcolor{negligible}0.05 & \cellcolor{negligible}-0.03 & \cellcolor{negligible}-0.10 \\ \hline
\security & \cellcolor{negligible}-0.09 & \cellcolor{negligible}-0.10 & \cellcolor{negligible}-0.10 & \cellcolor{negligible}-0.03 & \cellcolor{negligible}0.16 & \cellcolor{negligible}-0.08 & \cellcolor{negligible}-0.03 & \cellcolor{negligible}-0.06 & \cellcolor{negligible}-0.10 & \cellcolor{negligible}-0.05 & \cellcolor{negligible}-0.06 & \cellcolor{vstrong_pos}\color{white}1.00 & \cellcolor{negligible}-0.06 & \cellcolor{weak_pos}0.20 & \cellcolor{negligible}0.01 \\ \hline
\economics & \cellcolor{negligible}-0.07 & \cellcolor{negligible}0.08 & \cellcolor{negligible}-0.08 & \cellcolor{negligible}-0.11 & \cellcolor{negligible}-0.07 & \cellcolor{negligible}0.03 & \cellcolor{negligible}0.10 & \cellcolor{negligible}-0.05 & \cellcolor{weak_pos}0.22 & \cellcolor{negligible}0.14 & \cellcolor{negligible}0.05 & \cellcolor{negligible}-0.06 & \cellcolor{vstrong_pos}\color{white}1.00 & \cellcolor{negligible}-0.08 & \cellcolor{negligible}-0.08 \\ \hline
\verif & \cellcolor{negligible}-0.08 & \cellcolor{negligible}-0.14 & \cellcolor{negligible}-0.14 & \cellcolor{weak_neg}-0.21 & \cellcolor{weak_pos}0.22 & \cellcolor{negligible}-0.11 & \cellcolor{negligible}-0.13 & \cellcolor{negligible}-0.08 & \cellcolor{negligible}-0.14 & \cellcolor{negligible}-0.10 & \cellcolor{negligible}-0.03 & \cellcolor{weak_pos}0.20 & \cellcolor{negligible}-0.08 & \cellcolor{vstrong_pos}\color{white}1.00 & \cellcolor{negligible}0.08 \\ \hline
\pldesign & \cellcolor{negligible}-0.12 & \cellcolor{negligible}-0.12 & \cellcolor{negligible}-0.13 & \cellcolor{negligible}-0.08 & \cellcolor{negligible}-0.12 & \cellcolor{negligible}-0.10 & \cellcolor{negligible}-0.12 & \cellcolor{negligible}-0.08 & \cellcolor{negligible}-0.13 & \cellcolor{negligible}-0.13 & \cellcolor{negligible}-0.10 & \cellcolor{negligible}0.01 & \cellcolor{negligible}-0.08 & \cellcolor{negligible}0.08 & \cellcolor{vstrong_pos}\color{white}1.00 \\ \hline
\bottomrule
\end{tabular}
\end{adjustbox}
\end{table}

\end{document}